\documentclass[
reprint,amsmath,amssymb,
aps]{revtex4-2}
\usepackage{graphicx}% Include figure files
\usepackage{dcolumn}% Align table columns on decimal point
\usepackage{lipsum}% Este paquete es para agregar textos en latin usando \lipsum[1] 
\usepackage{bm}% bold math
\usepackage[colorlinks=true, allcolors=blue]{hyperref}
\usepackage[utf8]{inputenc}
\usepackage{graphicx} 
\usepackage{wrapfig} 
\usepackage[font=footnotesize]{caption}
\usepackage{amssymb,amsmath,amsthm}
\usepackage{enumerate}
\usepackage{anysize}
\usepackage{array}
\usepackage{float}
\usepackage{url}
\usepackage{soul}
\usepackage{physics}
\usepackage{mathtools}
\usepackage[usenames,dvipsnames,svgnames,table]{xcolor}
\usepackage[colorlinks=true, allcolors=blue]{hyperref}
\usepackage{geometry}
\usepackage{caption}% Este paquete tanto como el de subcaption se agregan para poder poner más de una figura en una imágen
\usepackage{subcaption}% Este paquete tanto como el de caption se agregan para poder poner más de una figura en una imágen
\usepackage{fancyhdr}
\usepackage{multirow} % Este paquete permite combinar filas en las tablas
\usepackage[mathscr]{euscript} 
{\par\medskip\noindent\minipage{\linewidth}}
{\endminipage\par\medskip}

\begin{document}
	
	\preprint{APS/123-QED}
	
	\title{Quantum corrections to cosmic perturbations for a bouncing background}
	
	\author{Hector Hernandez-Hernandez}%
	\email{hhernandez@uach.mx}
	\affiliation{Universidad Autonoma de Chihuahua, Facultad de Ingenieria, \\
		Nuevo Campus Universitario, Chihuahua 31125, México
	}
	\affiliation{
		Universidad Autonoma Metropolitana Cuajimalpa, Departamento de Matematicas Aplicadas y Sistemas, Vasco de Quiroga 4871, Ciudad de México 05348, México}
	\author{Hugo A. Morales-Técotl}
	\email{hugo@xanum.uam.mx}
	\author{Gustavo Sanchez-Herrera}
	\email{cbi2233805002@xanum.uam.mx}
	\affiliation{Universidad Autonoma Metropolitana Unidad Iztapalapa, San Rafael Atlixco 186, CP 09340,\\ 
		Ciudad de México, México.
	}

	\date{\today}% It is always \today, today,
	%  but any date may be explicitly specified
	
	\begin{abstract}
		We compute second-order quantum dispersions and correlations corrections, within an effective framework describing a single scalar perturbation mode propagating on a bouncing Loop Quantum Cosmology (LQC) background. Using an effective quantization approach in which quantum moments extend the classical phase space as new dynamical degrees of freedom, and incorporating the cosmic bounce through holonomy corrections in the $\mu_0$ scheme, we derive a coupled set of effective equations of motion for the expectation values and second-order quantum moments of both the gravitational and scalar sectors evolving with respect to a clock scalar field. Within the test-field approximation and for a vanishing scalar potential, the quantum moment equations reduce to a third-order ordinary differential equation for the mean squared deviation $G^{vv}$ of the Mukhanov--Sasaki variable in a de~Sitter background with LQC bounce. Treating the effect of bounce as a perturbation of the solution, we construct the corresponding correction to the dimensionless curvature power spectrum. The leading correction is suppressed by the sixth power of the Planck length producing a scale-dependent enhancement $\delta\mathcal{P}_\mathcal{R}\propto (k\,\ell_\mathrm{Pl})^6$ that modifies the spectral index by $\delta n_s \sim 6(k\,\ell_\mathrm{Pl})^6\ll 1$ for all cosmologically observable modes, in full consistency with current observational constraints. Numerical evolution of the full coupled system reveals a conditional ultraviolet regularization of the bounce-induced spectrum: the gravitational quantum moments generate a damping mechanism that suppresses the scalar perturbation amplitude after the bounce. Including cross-sector quantum correlations amplifies perturbation modes and introduces numerical instabilities at high wavenumbers, signaling the limits of the second-order truncation in the ultraviolet. \\
		
	\end{abstract}
	
	%\keywords{Suggested keywords}%Use showkeys class option if keyword                              %display desired
	\maketitle
	%\tableofcontents
	
	\section{INTRODUCTION}
	
	The theory of cosmological perturbations provides the bridge
	connecting the very early Universe with the large-scale structures observed
	today~\cite{mukhanov1992theory}. Among its most remarkable predictions are the
	anisotropies in the Cosmic Microwave Background
	(CMB)~\cite{sachs1967perturbations}, the statistical characterization of primordial
	perturbations and their role in structure
	formation~\cite{peebles1970primeval}, and the quantum generation of an
	almost scale-invariant spectrum of fluctuations during
	inflation~\cite{mukhanov1981quantum}. These predictions have been confirmed
	with increasing precision by a succession of observational
	programs~\cite{smoot1992structure, spergel2003first, ade2014planck, tegmark2004three, eisenstein2005detection}.
	
	In the standard formulation, small first-order fluctuations of the
	gravitational and matter sectors evolve on a fixed classical
	Friedmann--Lema\^{i}tre--Robertson--Walker (FLRW) background, and
	Einstein's field equations decouple into independent scalar, vector,
	and tensor sectors, each with distinct physical
	implications~\cite{bardeen1980gauge, mukhanov1992theory}.
	
	Only the perturbations
	are quantized in this framework, while the background remains
	classical. At the very early epochs where quantum effects are
	dominant, however, it is natural to expect that quantum corrections
	affect not only the perturbations but also the background dynamics
	itself, together with non-negligible backreaction effects between the
	two sectors.
	
	A compelling consequence of quantizing the gravitational background
	using Loop Quantum Cosmology (LQC) is the generic replacement of the
	initial singularity by a quantum
	bounce~\cite{bojowald2001absence, bojowald2001dynamical, bojowald2002isotropic, ashtekar2006quantum}.
	For more general cosmological scenarios---including models with matter
	fields or reduced symmetry such as anisotropies and
	inhomogeneities---where closed-form solutions are unavailable,
	effective quantization methods provide a powerful and systematic tool
	for resolving the initial singularity and extracting physical
	predictions~\cite{bojowald2007effective, hernandez2024singularity}. Among
	these, the approach based on quantum moments, statistical dispersions
	and correlations of quantum observables treated as new dynamical
	degrees of freedom extending the classical phase
	space~\cite{bojowald2006effective, bojowald2012quantum}, has proven effective
	across a broad class of quantum systems, from isotropic and anisotropic
	cosmologies to quantum-mechanical
	models~\cite{valdez2025effective, hernandez2023semiclassical, hernandez2024singularity, bojowald2024chaotic, brizuela2019moment, bojowald2021canonical}.
	
	Combining a quantum bounce with a consistent quantum treatment of
	cosmological perturbations is a central open problem. Significant
	progress has been made within LQC and related frameworks, where
	quantum gravitational effects produce a nonsingular cosmic
	evolution~\cite{bojowald2009gauge, agullo2013extension, agullo2012quantum}. In particular,
	the hybrid quantization approach ~\cite{agullo2013extension, agullo2012quantum} extends the quantum theory of
	cosmological perturbations to the Planck era by treating the
	background as a quantum geometry through a fully nonperturbative
	quantization of the homogeneous sector via LQC, while the
	perturbations are promoted to quantum fields propagating on that
	dressed quantum background. 
	This approach has produced detailed predictions for the primordial
	power spectrum and non-Gaussianities in the pre-inflationary
	bounce era, including
	explicit confrontations with CMB observations that have established
	the conditions under which pre-bounce dynamics can leave observable
	imprints on the power spectrum~\cite{agullo2013pre,ashtekar2021cosmic}.
	More recently, the dependence of the primordial power spectrum
	on the choice of the regularization scheme within LQC has been
	systematically investigated in Ref.~\cite{kowalczyk2025primodial}, showing that
	while the detailed spectral features depend on the specific scheme
	employed, the qualitative structure of the bounce-induced corrections
	is robust across different regularizations.
	
	The present work follows a different and complementary strategy. Rather
	than separating the background and perturbative sectors into a
	nonperturbative quantum geometry and a quantum field theory on it,
	we treat both the background and the perturbation mode within
	the effective quantum moments formalism~\cite{bojowald2006effective, bojowald2023quasiclassical}. In this approach, the full
	quantum system---background plus scalar perturbation---is replaced by
	a semiclassical system on an extended phase space in which quantum
	dispersions and correlations appear as additional dynamical variables.
	Quantum backreaction between the two sectors is then encoded
	perturbatively through the coupling of these quantum moments, without
	requiring the construction of a Hilbert space for the background
	geometry. The method is particularly well-suited to situations where
	the full quantum dynamics is intractable and closed-form solutions
	are unavailable~\cite{bojowald2012quantum}.
	
	Within the quantum moments framework, second-order quantum corrections to
	scalar cosmological perturbations in a singular de~Sitter background were
	computed in \cite{brizuela2019moment}, deriving a third-order differential
	equation for the mean squared deviation $G^{vv}$ of the Mukhanov--Sasaki
	variable and constructing the corresponding correction to the primordial
	power spectrum. The
	present paper extends that analysis to a nonsingular, bouncing
	background: we incorporate LQC holonomy corrections in the $\mu_0$
	scheme into the effective Hamiltonian, derive the full second-order
	effective dynamics for the coupled gravitational-perturbation system,
	and obtain both an analytical correction to the power spectrum and
	a complete numerical evolution of the system through the bounce.
	This is the first time the quantum
	moments approach has been applied to scalar cosmological perturbations
	propagating on a LQC bouncing background.
	
	It should be noted that several bounce models that modify the
	primordial power spectrum or non-Gaussianities have been found to be
	in tension with Planck
	data~\cite{van2023constraining}. However, the constraints of that work
	apply primarily to models whose bounce-induced corrections are
	significant at cosmologically observable scales. As we show below, the
	LQC corrections computed here are suppressed by the sixth power of the
	Planck length and are therefore negligible at all scales accessible
	to current or foreseeable observations. Our model is therefore consistent with existing observational data,
	since LQC corrections are intrinsically negligible at all
	cosmologically accessible scales. The physical interest of the
	calculation lies in establishing the theoretical structure of these
	quantum corrections within the moments framework.
	
	The article is organized as follows. In Section~\ref{sec:perturbations}
	we introduce the Hamiltonian formulation of the scalar sector of
	cosmological perturbations in a flat FLRW background. In
	Section~\ref{sec:LQC} we introduce the LQC holonomy correction and
	the formalism of effective quantum mechanics. In
	Section~\ref{sec:dynamics} we derive the second-order effective
	dynamics and reduce it to the case of quantum scalar perturbations
	in a classical de~Sitter spacetime with vanishing potential for the scalar field.
	In Section~\ref{sec:numerical} we present the numerical evolution
	of the full coupled system. Finally, in
	Section~\ref{sec:conclusions} we discuss our results and outline
	directions for future work.
	
	\section{Scalar cosmological perturbations in a flat FLRW background}\label{sec:perturbations}
	
	In Lagrangian formalism, the action corresponding to the gravitational and scalar matter fields is
	\begin{eqnarray}\label{EH&ScalarAction}
		S &=&  \int \sqrt{-g} d^{4}x \left[- \frac{1}{6 l^{2}} R + \frac{1}{2}\phi_{,\alpha}\phi^{,\alpha}-U(\phi) \right],
	\end{eqnarray}
	where $l^2 = 8\pi G/3$, $R$ is the Ricci scalar,
	$g = \det(g_{\mu\nu})$ is the determinant of the spacetime metric,
	$\phi$ is the scalar field, and $U(\phi)$ is its potential. 
	
	In the standard theory of cosmological perturbations, the action~(\ref{EH&ScalarAction})
	is expanded to second order in small fluctuations of the gravitational
	and matter sectors, yielding independent scalar, vector, and tensor
	contributions, each with distinct physical implications~\cite{mukhanov1992theory}.
	Scalar perturbations are of particular interest due to their direct
	connection with the formation of large-scale structure. At linear order
	the three sectors decouple and can be treated
	separately.
	
	Expanding~(\ref{EH&ScalarAction}) to second order in scalar perturbations gives
	the Mukhanov--Sasaki action~\cite{mukhanov1992theory}
	\begin{eqnarray}\label{SecondOrderAction}
		S_{2} &=&  \frac{1}{2} \int d^{4}x \left(v'^{2}-v_{,i}v_{,i}+ \frac{z''}{z} v^{2}\right),
	\end{eqnarray}
	where $v = a\bigl[\delta\phi + (\phi'/\mathcal{H})\Psi\bigr]$
	is the gauge-invariant Mukhanov--Sasaki variable,
	$a$ is the scale factor, $\mathcal{H} = a'/a$ is the conformal Hubble
	parameter, $\delta\phi$ and $\Psi$ are the scalar matter and curvature
	perturbations respectively, and primes denote derivatives with respect to
	conformal time $\eta$, namely, $d\eta = a^{-1}dt$. The function $z''/z$ plays the role of an
	effective time-dependent mass that couples the perturbation to the
	background, with $z = a\phi'/\mathcal{H}$~\cite{mukhanov1992theory}.
	In practice, obtaining~(\ref{SecondOrderAction}) involves substantial algebraic work;
	an equivalent route is to derive Bardeen's equation for the
	Newtonian potential~\cite{bardeen1980gauge} and then perform appropriate
	changes of variables.
	
	The evolution of scalar cosmological perturbations is governed by the
	Mukhanov--Sasaki equation, obtained directly from~(\ref{SecondOrderAction}),
	\begin{eqnarray}
		v''_{k} + \omega^{2}v_{k}=0,
		\label{eq:MS}
	\end{eqnarray}
	with the time-dependent effective frequency $\omega^2 = k^2 - z''/z$.
	The dimensionless curvature power spectrum is related to the mode
	function $v_k$ by~\cite{baumann2022cosmology} 
	\begin{eqnarray}\label{PowerSpectrum}
		\mathcal{P}_{\mathcal{R}}&=&  \frac{k^{3}}{2 \pi^{2}}\frac{|v_{k}|^{2}}{z^{2}}.
	\end{eqnarray}

	Defining the canonically conjugate momentum $\pi = \partial\mathcal{L}/\partial v' = v'$
	and performing a Legendre transformation, the Hamiltonian corresponding
	to~(\ref{SecondOrderAction}) is
	\begin{eqnarray}\label{SecondOrderHamiltonian}
		H_{2} &=&  \frac{1}{2} \int d^{3}x \left(\pi^{2} +v_{,i}v_{,i} - \frac{z''}{z} v^{2}\right).
	\end{eqnarray}
	In the standard treatment, cosmic perturbations evolve on a given
	background and $z$ depends only on background quantities, so the
	effective mass $z''/z$ is a prescribed function of conformal time
	that remains unchanged by the Legendre transformation.
	For a more general description in which the background and perturbations
	are treated on an equal footing, however, a complete canonical expression
	for $z''/z$ is required. Through canonical transformations it can be
	shown~\cite{pinho2007scalar} that the Hamiltonian corresponding
	to~(\ref{EH&ScalarAction}) and~(\ref{SecondOrderAction}) in a flat FLRW background is
	\begin{eqnarray}\label{GravitationalMatterScalarHamiltonian}
		H  &=& N \left[-\frac{l^{2}P_{a}^{2}}{4 a \mathcal{V}} + \frac{P_{\phi}^{2}}{2 a^{3}\mathcal{V}} + \frac{a^{3}\mathcal{V} U}{2} \right. \nonumber \\
		&& \left.+ \frac{1}{2a}\int dx^{3} \left(\pi^{2} +  v^{,i}v_{,i} - \zeta v^{2}\right)\right],
	\end{eqnarray}
	where $N$ is the lapse function; $P_a = -2a\dot{a}/l^2$ and
	$P_\phi = a^3\dot{\phi}$ are the canonical momenta satisfying
	$\{a, P_a\} = l^2$ and $\{\phi, P_\phi\} = 1$; $\mathcal{V}$ is the
	fiducial spatial volume introduced to regulate spatial integrations;
	and dots denote derivatives with respect to cosmic time $t$.
	The time-dependent function $\zeta$ is given by~\cite{pinho2007scalar}
	\begin{eqnarray}
		\zeta &=&-\left(\frac{15 l^{2}P_{\phi}^{2}}{2 a^{4}\mathcal{V}^{2}}+\frac{a^{2} U_{,\phi \phi}}{2}-\frac{3l^{2}a^{2}U }{4}+\frac{9 U P_{\phi}^{2}}{2P_{a}^{2}} \right.\nonumber \\
		&&\left. -\frac{l^{4}P_{a}^{2}}{8 a^{2}\mathcal{V}^{2}}-\frac{27 P_{\phi}^{4}}{2 a^{6}P_{a}^{2}\mathcal{V}^{2}}-\frac{6 aP_{\phi}U_{,\phi}}{P_{a}}\right). 
	\end{eqnarray}
	Here $U_{,\phi} = dU/d\varphi$. Using the background equations of motion, it can be
	verified that $\zeta$ is equivalent to $z''/z$
	in~(\ref{SecondOrderAction}) (see  Ref.\,\cite{pinho2007scalar}.)
	
	Since our goal is to obtain an effective dynamics coupled to a
	nonsingular background, it is sufficient to consider a single
	perturbation mode in~(\ref{GravitationalMatterScalarHamiltonian}). This restriction limits
	the analysis to the linear-order physics and precludes the study
	of non-Gaussianities or the bispectrum, which would require
	retaining multiple modes, which we leave for future work. Taking the Fourier transform of $v$ in
	flat space
	\begin{eqnarray}
		v &=& \frac{1}{(2\pi)^{3/2}} \int d^{3}k v_{k}e^{-ik\cdot x},
	\end{eqnarray}
	with $k$ having units $[k]=\text{Length}^{-1}$, the Hamiltonian for each scalar mode $k$ becomes
	\begin{eqnarray}\label{GravitationalMatterSingleScalarHamiltonian}
		H_{\text{M}}  &=&  N \left[-\frac{l^{2}P_{a}^{2}}{4 a \mathcal{V}} + \frac{P_{\phi}^{2}}{2 a^{3}\mathcal{V}} + \frac{a^{3}\mathcal{V} U}{2}  + \frac{1}{2a}\left(\pi_{k}^{2}  + \omega_{s}^{2} v_{k}^{2}\right)\right]. \nonumber \\
	\end{eqnarray}
	Here, $v_k$ and $\pi_k$ are the Fourier coefficients of $v$ and $\pi$
	respectively, and the scalar mode behaves as a harmonic oscillator with
	time-dependent frequency $\omega_s^2(k,\zeta) = k^2 - \zeta$.
	
	The Hamiltonian~(\ref{GravitationalMatterSingleScalarHamiltonian}) can be deparametrized by using the scalar field $\phi$ as internal time and setting the lapse function
	$N = 1$. Applying the constraint $H_M \approx 0$, expanding the
	frequency $\omega_s^2 = k^2 - \zeta$, and rearranging terms, one obtains a quartic
	polynomial equation for $P_\phi$,
	\begin{eqnarray}\label{FourPolynomicalEquation}
		\lambda P_{\phi}^{4} +\mu \frac{P_{\phi}^{2}}{2} +\nu P_{\phi}+ \xi \approx 0,
	\end{eqnarray}
	where the coefficients are 	
	\begin{eqnarray}\label{Functionsalphadelta}
		\lambda &=& -\frac{27v_{k}^{2}}{4a^{7}P_{a}^{2}}, \nonumber \\
		\mu &=&  \frac{1}{a}\left(\frac{15 l^{2}v_{k}^{2}}{2 a^{4}} +  \frac{1}{a^{2}} +\frac{9U v_{k}^{2}}{2P_{a}^{2}}\right), \nonumber \\
		\nu &=&  - \frac{3U_{,\phi}v_{k}^{2}}{P_{a}}, \nonumber \\
		\xi &=& \left(-\frac{l^{2}P_{a}^{2}}{4a} +\frac{a^{3}U}{2} +\frac{1}{2a}\left(\pi_{k}^{2} + k^{2}v_{k}^{2}\right)\right. \nonumber \\
		&&\left. + \frac{a^{2}U_{,\phi, \phi}v_{k}^{2}}{4 a} -\frac{3l^{2}Ua^{2}v_{k}^{2}}{8a}- \frac{l^{4}P_{a}^{2}v_{k}^{2}}{16 a^{3}} \right).
	\end{eqnarray}
	For a constant potential $U$, the term $a^3 \mathcal{V} U/2$
	in~(\ref{GravitationalMatterScalarHamiltonian}) becomes subdominant relative to the kinetic
	contribution as $a\to 0$, i.e., near the initial classical singularity. In
	this regime the coefficient $\nu$ vanishes and~(\ref{FourPolynomicalEquation})
	reduces to
	\begin{eqnarray}\label{CuadraticEquation}
		\lambda P_{\phi}^{4} +\mu \frac{P_{\phi}^{2}}{2}+ \xi \approx 0,
	\end{eqnarray}
	with general solutions 
	\begin{eqnarray}\label{4GeneralSolutions}
		P_{\phi_{1,2}}&=&\pm\sqrt{-\frac{1}{4\lambda}\left(-\mu + \sqrt{\mu^{2} - 16\lambda\xi}\right)}, \nonumber \\
		P_{\phi_{3,4}} &=& \pm\sqrt{-\frac{1}{4\lambda}\left(\mu + \sqrt{\mu^{2} - 16\lambda\xi}\right)}.
	\end{eqnarray}
	The sign of the outer square root in~(\ref{4GeneralSolutions}) determines
	the direction of time~\cite{bojowald2007large}; requiring positive time
	evolution eliminates the negative branches. Taylor-expanding the
	inner square root to first order in $\lambda\xi/\mu^2$, 
	valid when $|\lambda\xi/\mu^2| \ll 1$ 
	(which follows from the small-amplitude condition $v_k/a\ll 1$ assumed throughout),
	yields
	\begin{eqnarray}\label{ReducedSolutions}
		\frac{P_{\phi_{1}}}{\sqrt{2}}&=& \sqrt{- \frac{\xi}{\mu}}, \hspace{0.5cm} \frac{P_{\phi_{3}}}{\sqrt{2}}=\sqrt{-  \frac{1}{4\lambda}\left(\mu -4\frac{\lambda\xi}{\mu}\right)},
	\end{eqnarray}
	These two solutions can be identified as deparametrized Hamiltonians
	$\mathcal{H}$ and $\mathcal{H}'$, respectively, whose equations of
	motion are generated by derivatives with respect to $\phi$.
	
	Substituting~(\ref{Functionsalphadelta}) into~(\ref{ReducedSolutions}) and
	employing the small-amplitude condition $v_k/a \ll 1$, the two
	deparametrized Hamiltonians for constant potential $U$ become
	\begin{eqnarray}\label{ClassicalHamiltonian}
		\mathcal{H} &:=& - \frac{P_{\phi_{1}}}{\sqrt{2}} 
		\approx -
		\sqrt{  \frac{l^{2}a^{2}P_{a}^{2}}{4} - \frac{a^{6}U}{2}  - a^{2}H_{\text{s}} }, \nonumber \\
		\mathcal{H}'&:=&- \frac{P_{\phi_{3}}}{\sqrt{2}}
		\approx    \frac{1}{2\sqrt{27} } \frac{a^{2}P_{a}}{v_{k}} ,
	\end{eqnarray}	
	where $H_s \equiv (\pi_k^2 + k^2 v_k^2)/2$ is the
	Hamiltonian of a single scalar mode with frequency $k$.
	The quantity $v_k/a$ describes the physical amplitude of the
	perturbation: since $z = a\dot{\phi}/H$ and
	$\mathcal{P}_\mathcal{R} \propto (|v_k|/a)^2$, the condition
	$v_k/a \ll 1$ ensures that the mode remains a small perturbation
	on the background.
	
	In this work we focus on $\mathcal{H}$. The Hamiltonian $\mathcal{H}'$
	is discarded because, from Eq.~(\ref{ClassicalHamiltonian}),
	$\mathcal{H}' \approx a^2 P_a / (2\sqrt{27}\, v_k)$,
	which diverges as $v_k \to 0$. This divergence reflects the fact that
	$\mathcal{H}'$ describes a sector in which the scalar mode amplitude
	vanishes while the gravitational contribution $a^2 P_a$ remains finite,
	a regime that is physically inconsistent with treating $v_k$ as a
	propagating perturbation. Furthermore, $\mathcal{H}'$ does not reduce
	to the standard Hamiltonian for scalar perturbations on a cosmological
	background in the limit $v_k/a \ll 1$, unlike $\mathcal{H}$, which
	correctly reproduces that limit through the identification of
	$H_s = (\pi_k^2 + k^2 v_k^2)/2$ as the mode energy.
	
	\section{Loop quantum cosmology correction and Effective quantum mechanics}\label{sec:LQC}
	
	\subsection{Loop quantum cosmology correction}
	The difficulty of Loop Quantum Gravity (LQG) when applied to
	realistic cosmological scenarios has motivated the development of a
	reduced framework obtained by imposing the cosmological principle,
	known as Loop Quantum Cosmology (LQC). This approach inherits the
	central features of LQG while providing a tractable setting for
	studying the early Universe. LQC is most conveniently formulated
	within a canonical framework based on the Arnowitt--Deser--Misner
	(ADM) decomposition of general relativity, expressed in terms of the
	Ashtekar--Barbero variables~\cite{ashtekar1986new}. This formalism
	requires the introduction of triads $e^a{}_i$ and co-triads
	$e^i{}_a$, satisfying $q_{ab}e^a{}_i e^b{}_j = \delta_{ij}$,
	$e^a{}_i e^b{}_j = \delta^i_j$, and $e^a{}_i e^b{}_j = \delta^b_a$,
	where $i,j,k,\ldots$ and $a,b,c,\ldots$ denote internal and spatial
	indices, respectively. From these one constructs the densitized triad
	$E^a{}_i \equiv \sqrt{q}\,e^a{}_i$ and the extrinsic curvature
	$K^i{}_a \equiv K_{ab}e^{bi}$.
	
	The canonical variables are the Ashtekar--Barbero connection
	$A^i{}_a = \Gamma^i{}_a + \gamma K^i{}_a$ and the densitized triad
	$E^a{}_i$, satisfying the Poisson
	bracket~\cite{bojowald2010canonical}
	\begin{equation}
		\{A^j{}_a(x),\, E^b{}_k(y)\}
		= 8\pi G\gamma\,\delta^b_a\,\delta^j_k\,\delta^3(x,y),
	\end{equation}
	where $\Gamma^i{}_a$ is the spin connection and $\gamma$ is the
	Barbero--Immirzi parameter. The gravitational Hamiltonian reads
	\begin{eqnarray}\label{FullGravitationalConstraint}
		H_{\text{g}} [A_{a}^{i},E_{j}^{b}]&=& \int d^{3}x ( N\mathcal{C}_{\text{g}}+N^{a}\mathcal{C}_{a}^{\text{g}}-\Lambda^{i}\mathcal{G}_{i}),
	\end{eqnarray}
	where $\mathcal{C}_g$, $\mathcal{C}^g_a$, and $\mathcal{G}_i$ are
	the Hamiltonian, diffeomorphism, and Gauss constraints,
	respectively. These encode the dynamics and ensure invariance under
	spacetime diffeomorphisms and local $SU(2)$ rotations. They are
	defined as
	\begin{eqnarray}\label{GravityConstraints}
		\mathcal{C}_{\text{g}} &=&-4\pi G\gamma^{2}\epsilon^{ij}_{\hspace{0.1cm}k} \frac{P_{i}^{a}P_{j}^{b}}{\sqrt{q}}\left(\mathcal{F}_{ab}^{k}+(1+\gamma^{2})\epsilon^{k}_{\hspace{0.1cm}mn}K_{a}^{m}K_{b}^{n}\right) \approx 0 \nonumber \\
		\mathcal{C}_{a}^{\text{g}} &=& \mathcal{F}^{i}_{ab} E^{b}_{i} \approx 0, \nonumber \\
		\mathcal{G}_{i} &=& \mathcal{D}_{b}^{(A)}E_{j}^{b}\approx 0, 
	\end{eqnarray}
	with $P_{i}^{a}\equiv E_{i}^{a}/8 \pi G \gamma $ and $\mathcal{F}^{l}_{ab} = F_{ab}^{l} + 2 \gamma \mathcal{D}_{[a}K_{b]}^{l}-\gamma^{2}\epsilon^{l}_{\hspace{0.1cm}jk}K_{a}^{j}K_{b}^{k}$ with the curvature connection $F^{l}_{ab}:= 2 \partial_{[a}\Gamma_{b]}^{l}-\epsilon^{l}_{\hspace{0.1cm}jk}\Gamma_{[a}^{j}\Gamma_{b]}^{k}$.
	
	For a homogeneous and isotropic universe, the full nontrivial
	dynamics is encoded entirely in the Hamiltonian
	constraint~\cite{bojowald2000loop}. The diffeomorphism and Gauss
	constraints are automatically satisfied by the symmetries of
	spacetime. Under the cosmological principle, the infinite-dimensional
	phase space of general relativity reduces to a finite-dimensional
	one~\cite{ashtekar2003mathematical}, and the Ashtekar connection and densitized
	triad take the form
	\begin{eqnarray}
		A_{a}^{i} = c V_{0}^{-1/3}  {}^{\circ} e_{a}^{i}, \hspace{0.5cm} E_{i}^{a} = p V_{0}^{-2/3}\sqrt{{}^{\circ} q}{}^{\circ} e_{i}^{a},
	\end{eqnarray}
	where $c(t) = \gamma\dot{a}$ and $|p(t)| = a^2$ are the reduced
	canonical variables, $V_0$ is the fiducial cell volume, and
	${}^\circ\!e^i{}_a$, ${}^\circ\!q$ are fiducial quantities
	introduced to avoid infrared divergences in the spatially flat FLRW
	model. Their Poisson bracket is~\cite{bojowald2007effective}
	\begin{eqnarray}
		\lbrace c, p \rbrace &=& \gamma l^{2}.
	\end{eqnarray}
	In terms of $c$ and $p$, the canonical momentum conjugate to the
	scale factor is
	\begin{eqnarray} \label{eq:Pa}
		P_{a}^{2} &=&  \frac{4a^{2}\dot{a}^{2}}{l^{4}}  = \frac{4 pc^{2}}{\gamma^{2}l^{4}}. 
	\end{eqnarray}
	In LQC, the Ashtekar connection is not a well-defined operator on
	the kinematic Hilbert space. Instead, the curvature of the
	connection is expressed through holonomies---parallel transports
	of $A^i{}_a$ along loops of coordinate area $\mu^2$. The leading
	quantum correction to the classical connection replaces $c$ with
	$\sin(\mu c)/\mu$, which reduces to $c$ as $\mu\to 0$ and
	introduces a maximum energy density $\rho_c \sim \rho_{\mathrm{Pl}}$
	at which the classical singularity is replaced by a
	bounce~\cite{ashtekar2006quantum,bojowald2001absence}. In the $\mu_0$ scheme
	adopted here, the discretization parameter $\mu$ is taken to be
	constant~\cite{bojowald2007effective}.
	
	To incorporate this holonomy correction and simultaneously linearize
	the Hamiltonian expression, we introduce the non-canonical
	complex variables~\cite{bojowald2007effective}
	\begin{equation}
		J := p\,e^{ic}, \qquad \bar{J} := p\,e^{-ic}.
		\label{eq:JJbar}
	\end{equation}
	More precisely, these are $\mu$-dependent: $J_\mu = \mu^{-1}p\,
	e^{i\mu c}$ and $\bar{J}_\mu = \mu^{-1}p\,e^{-i\mu c}$, where the
	holonomy correction $c\to\sin(\mu c)/\mu$ is encoded in the
	imaginary exponent. For notational simplicity we set $\mu=1$
	throughout the remainder of this section; the $\mu$-dependence is
	restored explicitly in Section~\ref{subsec:classicallimit} when we
	discuss the classical limit. The variables $J$ and $\bar{J}$
	satisfy the Poisson algebra~\cite{bojowald2007effective}
	\begin{eqnarray}\label{nonCanonicalVariables}
		\lbrace p , J \rbrace = -i \gamma l^{2} J,  \hspace{0.3cm}
		\lbrace p , \bar{J} \rbrace = i \gamma l^{2}\bar{J}, \hspace{0.3cm} 
		\lbrace J , \bar{J} \rbrace = 2 i \gamma l^{2} p. \nonumber \\  
	\end{eqnarray}
	Using ~(\ref{eq:Pa}) to  rewrite $\mathcal{H}$ in~(\ref{ClassicalHamiltonian}), and expressing the result in terms of $J$
	and $\bar{J}$, we obtain the Hamiltonian
	\begin{equation}
		\mathcal{H}_{\text{bs}} =- \sqrt{-\frac{1}{4\gamma^{2}l^{2}}\left(J-\bar{J} \right)^{2} - \frac{p^{3}U}{2}-p H_{\text{s}} },
		\label{eq:Hbs_full}
	\end{equation}
	which encodes the cosmic bounce through the holonomy-corrected
	variables and is coupled to the scalar perturbation mode via
	$H_{\text{s}}$.
	
	The gravitational term $(J-\bar{J})^2/(4\gamma^2 l^2)$ dominates
	near the bounce, where the energy density approaches $\rho_c$.
	In the regime where the matter and perturbation contributions are
	small compared to the gravitational term---i.e., when
	$\rho_{\mathrm{matter}} + \rho_{\mathrm{pert}} \ll \rho_c$---we
	expand the square root to first order, obtaining
	\begin{eqnarray}\label{H1perturbated}
		\mathcal{H}_{\text{bs}} &\approx& - \frac{i l \gamma}{2} \left[\frac{1}{ l^2 \gamma^2}\left(J-\bar{J}\right) +  \frac{p^{3}U+ 2 H_{\text{s}} p}{(J-\bar{J})}  \right]. 
	\end{eqnarray}
	%
	
	%   \textcolor{red}{This approximation is most accurate near the bounce and       %  becomes less reliable as the system approaches the de~Sitter regime, where
		%  $\rho_{\mathrm{matter}} \sim \rho_c$.}
	
	\subsection{Effective quantum mechanics}
	
	Through a generalization of Ehrenfest's theorem, it is possible to
	formulate an effective description of a quantum system in which the
	dynamics is governed by an effective Hamiltonian
	$H_{\mathrm{eff}} = \langle\hat{H}\rangle$ that depends on the
	expectation values of observables as well as on their dispersions
	and correlations---the quantum moments. This approach yields
	physically relevant quantum information without requiring the
	solution of the Schr\"odinger equation, which is intractable for
	most complex quantum systems. The effective Hamiltonian is obtained
	through a Taylor expansion of $\langle\hat{H}\rangle$ around the
	expectation values of the canonical
	variables~\cite{bojowald2006effective}. Dynamical equations of motion
	follow from Dirac's prescription
	\begin{eqnarray}\label{OperatorsBracketComutation}
		\lbrace \langle \hat f \rangle, \langle \hat g \rangle \rbrace&=& \frac{1}{i \hbar}\langle [\hat f, \hat g] \rangle. 
	\end{eqnarray}  
	For a system with $k$ canonical pairs $(\hat{q}_i, \hat{p}_i)$,
	the quantum moments are defined as~\cite{bojowald2011high}
	\begin{align}\label{kpairsmoments}
		G^{a_{1},b_{1},\cdots,a_{k},b_{k}}&=  \langle  (\hat q_{1}-q_{1})^{a_{1}}\cdots (\hat q_{k}- q_{k})^{a_{k}}\nonumber \\ &\times(\hat p_{1}-p_{1} )^{b_{1}}\cdots(\hat p_{k}-p_{k} )^{b_{k}} \rangle_{\text{Weyl}}, 
	\end{align}
	where $q_i = \langle\hat{q}_i\rangle$, $p_i = \langle\hat{p}_i\rangle$,
	and the subscript denotes total Weyl (symmetric) ordering of the
	operator products~\cite{bojowald2011high}. For moments built
	from canonical variables, the Heisenberg uncertainty principle
	requires~\cite{bojowald2006effective}
	\begin{eqnarray}\label{MOmentsUncertanityRelation}
		G^{2,0}G^{0,2} - (G^{1,1})^{2} &\geq& \frac{\hbar^{2}}{4}.
	\end{eqnarray}
	The effective Hamiltonian is given by the formal series
	\begin{align}\label{GeneralKhamiltonian}
		H_{\text{eff}}&= \sum_{a_{1},b_{1} \ldots a_{k},b_{k}}^{\infty}
		\frac{G^{a_{1},b_{1},\cdots,a_{k},b_{k}}}{a_{1}!b_{1}!\cdots a_{k}!b_{k}!}\frac{\partial^{a_{1}+b_{1}+\cdots +a_{k}+b_{k}}H_{\text{cl}}}{\partial q_{1}^{a_{1}} \partial p_{1}^{b_{1}} \cdots \partial q_{k}^{a_{k}}  \partial p_{k  }^{b_{k}} }, \nonumber \\ &
	\end{align}
	where $H_{\mathrm{cl}}$ is the classical Hamiltonian. For a general
	system the sum in~(\ref{GeneralKhamiltonian}) is infinite, leading to an
	infinite hierarchy of equations of motion that cannot be solved in
	closed form. A finite, tractable system is obtained by
	truncating that expression at a chosen order in the quantum
	moments~\cite{bojowald2012quantum}. The truncation is justified by the
	$\hbar$-ordering of the moments: since $G^{a,b} \propto
	\hbar^{(a+b)/2}$, retaining only second-order moments (i.e.,
	$a+b = 2$) captures the leading quantum corrections beyond the
	classical behavior, with higher-order terms suppressed by
	additional powers of $\hbar$.
	
	To obtain solutions, initial conditions for the quantum moments must
	be specified. We employ a Gaussian state for the gravitational sector,
	\begin{align}\label{GaussP}
		\psi(c)&= \frac{1}{(2 \pi \sigma^{2})^{1/4}} 	\exp{-\frac{(c -\langle \hat{c} \rangle)^{2}}{4 \sigma^2}+ \frac{i  \langle \hat{p} \rangle}{\hbar}(c-\langle \hat{c} \rangle)}, 
	\end{align}
	where $\langle\hat{c}\rangle$ and $\langle\hat{p}\rangle$ fix the
	initial expectation values of the connection and its conjugate
	momentum, and $\sigma$ is the width of the wave packet. This choice
	is physically motivated: a Gaussian state is a minimum-uncertainty
	state and represents the closest quantum analogue to a classical
	configuration while still encoding nontrivial quantum fluctuations
	through its dispersions and correlations. The corresponding initial
	quantum moments are
	\begin{equation}
		G^{a,b}_0 = \begin{cases}
			\displaystyle
			2^{-(a+b)}\hbar^a \sigma^{b-a}\,\frac{a!\,b!}{(a/2)!\,(b/2)!},
			& a \text{ and } b \text{\ even,}\\[6pt]
			0, & \text{otherwise.}
		\end{cases}
		\label{GGC}
	\end{equation}
	The proportionality $G^{a,b} \propto \hbar^{(a+b)/2}$ confirms
	that the truncation of~(\ref{GeneralKhamiltonian}) at second order in quantum
	moments is controlled by powers of $\hbar$, with the truncation
	error entering at order $\hbar^{3/2}$ and beyond~\cite{bojowald2012quantum}.
	
	Note that in this work we perform the deparameterization at the classical level instead of first quantizing the full constraint and deparameterizing at the quantum level, as in the quantum formalism for constrained systems \cite{bojowald2009effective}. Therefore, no quantum moments of the scalar field appear. This approximation is justified since we consider an initially semiclassical Gaussian state with a sharply peaked wave function, such that the effective evolution closely follows the classical dynamics. Moreover, our effective analysis is restricted to the kinetic-energy-dominated regime around the bounce, where the potential $U$ can be neglected and $P_{\phi}^{2}=2\rho_{\mathrm{cr}}a_{\mathrm{Bounce}}^{6}$, with $\rho_{\mathrm{cr}}\approx0.41\,\rho_{\mathrm{Pl}}$. In this regime, the scalar field provides a suitable semiclassical clock since it evolves monotonically and its conjugate momentum $P_{\phi}$ is large and sharply peaked, keeping the associated quantum fluctuations small. The connection between our approach and the effective constraints formalism \cite{bojowald2009effective} can be reviewed in the Appendix \ref{AppendixEffectiveConstraints} 
	
	\section{Effective dynamics}\label{sec:dynamics}
	
	\subsection{Second order effective Dynamics}
	Following canonical quantization, the classical variables in
	Eq.~(\ref{H1perturbated}) are promoted to operators. Using the Poisson
	algebra~(\ref{nonCanonicalVariables}) and the standard quantization rule
	$\{\cdot,\cdot\} \to [\cdot,\cdot]/(i\hbar)$, the commutation
	relations between $\hat{p}$, $\hat{J}$, and $\hat{J}^\dagger$
	are~\cite{bojowald2007effective}
	\begin{align}
		[ \hat{p} , \hat{J}]  &= \gamma l^{2} \hbar \hat{J},  \hspace{1cm} [\hat{p} , \hat{J}^{\dagger}] = -\gamma l^{2}\hbar  \hat{J}^{\dagger}, \nonumber \\  
		& \hspace{0.1cm}  [ \hat{J} , \hat{J}^{\dagger} ] = -2 \gamma l^{2}\hbar(\hat{p}+\hbar/2). 
		\label{eq:commutators}
	\end{align}
	
	The quantum Hamiltonian operator corresponding
	to~(\ref{H1perturbated}) is
	\begin{align}\label{QH1perturbated}
		\hat{\mathcal{H}}_{\text{bs}} &\approx -\frac{i}{2 l\gamma}\left(\hat{J}-\hat{J}^{\dagger}\right) - \frac{i l \gamma}{2}\frac{\hat{p}^{3}U}{(\hat{J}-\hat{J}^{\dagger})} - \frac{i l \gamma\hat{p} \hat{H}_{\text{s}}}{(\hat{J}-\hat{J}^{\dagger})}. 
	\end{align}
	Expanding the expectation value
	$\langle\hat{\mathcal{H}}_{\mathrm{bs}}\rangle$ to second order in
	quantum moments using~(\ref{GeneralKhamiltonian}), we obtain the second-order
	effective Hamiltonian. The expansion proceeds by Taylor-expanding
	each operator product around the expectation values of $\hat{p}$,
	$\hat{J}$, $\hat{J}^\dagger$, $\hat{v}_k$, and $\hat{\pi}_k$, and
	collecting terms at each order in the moments; the zeroth-order
	terms reproduce the classical Hamiltonian~(\ref{GeneralKhamiltonian}), while the
	second-order terms generate all contributions proportional to
	$G^{JJ}$, $G^{Jp}$, $G^{vv}$, $G^{\pi\pi}$, and the
	cross-sector moments $G^{Jv}$, $G^{J\pi}$, $G^{\bar{J}v}$, $G^{\bar{J}\pi}$, $G^{pv}$, $G^{p\pi}$.
	The result is
	\begin{widetext}
		\small{
			\begin{eqnarray}\label{SecondOrderEffectiveHamiltonian}
				\mathcal{H}_{\text{bs}} ^{\text{eff}} &=&  -\frac{i}{2l\gamma}\Delta J - \frac{i l \gamma }{2} \frac{p^{3}U}{\Delta J} - i l\gamma H_{\text{s}} \frac{p}{\Delta J} -\frac{i l \gamma}{2} \frac{p}{\Delta J} \left(3 U G^{pp}+G^{\pi \pi}+ k^{2}G^{v v }\right) - \frac{i l \gamma}{2} \frac{ p}{(\Delta J)^{3}}\left(Up^{2}+ 2 H_{\text{s}}\right)\left(G^{JJ}+G^{\bar{J}\bar{J}}-2G^{J\bar{J}}\right) \nonumber  \\
				&&  + \frac{i l\gamma}{(\Delta J)^{2}}\left(\frac{3}{2}Up^{2}+H_{\text{s}}\right)\left(G^{Jp}-G^{\bar{J}p}\right)+ i l \gamma \frac{p}{(\Delta J)^{2}}\left[\pi_{k}\left(G^{J\pi}-G^{\bar{J}\pi}\right )+  k^{2} v_{k}(G^{Jv} +G^{\bar{J}v})\right]- \frac{i l \gamma}{\Delta J}\left(\pi_{k}G^{p\pi} + k^{2}v_{k}G^{pv}\right), \nonumber \\ 
			\end{eqnarray}
		}
	\end{widetext}
	where $H^\mathrm{eff}_\mathrm{bs} := \langle\hat{\mathcal{H}}_\mathrm{bs}\rangle$
	is the expectation value of the Hamiltonian operator,
	$\Delta J \equiv J - \bar{J}$, and we use the standard notation
	$\langle\hat{X}\rangle := X$ for expectation values. 
	
	The second-order effective equations of motion for the expectation
	values are
	\begin{eqnarray}\label{SecondOrderEOM1}
		Y_{,\phi} = F_{\text{2nd}}(Y, G^{ab}; \phi),
	\end{eqnarray}
	where 
	\begin{eqnarray}\label{componentsSecondOrderEOM1}
		Y &\equiv& \begin{pmatrix}
			J \\
			p \\
			v_{k} \\
			\pi_{k}
		\end{pmatrix}, \hspace{1cm }F_{\text{2nd}} \equiv \begin{pmatrix}
			F_J \\
			F_p \\
			F_{v_{k}} \\
			F_{\pi_{k}}
		\end{pmatrix}. 
	\end{eqnarray}
	The components of $F_\mathrm{2nd}$ are highly nonlinear functions
	of the expectation values and the second-order quantum moments.
	Similarly, the equations of motion for the second-order quantum
	moments are
	\begin{eqnarray}\label{SecondOrderEOM2}
		Z_{,\phi} = G_{\text{2nd}}(Y, G^{ab}; \phi),
	\end{eqnarray}
	where $Z$ and $G_{\text{2nd}}$  are given by 
	\begin{eqnarray}\label{componentsSecondOrderEOM2}
		Z &\equiv& \begin{pmatrix}
			G^{JJ} \\
			G^{pp} \\
			G^{vv} \\
			G^{\pi \pi}\\
			G^{J\bar{J}} \\
			G^{Jp} \\
			G^{Jv} \\
			G^{J \pi} \\
			G^{pv} \\
			G^{p \pi}\\
			G^{v \pi}
		\end{pmatrix}, \hspace{1cm }G_{\text{2nd}} \equiv \begin{pmatrix}
			G_1 \\
			G_2 \\
			G_3 \\
			G_4 \\
			G_5\\
			G_6 \\
			G_7 \\
			G_8 \\
			G_9 \\
			G_{10}\\
			G_{11} 
		\end{pmatrix}.
	\end{eqnarray}
	The explicit form of $F_\mathrm{2nd}$ and $G_\mathrm{2nd}$ is given
	in Appendix~\ref{Appendix}. Whenever an equation of motion
	admits a complex conjugate counterpart, the latter is obtained by
	direct conjugation; for example,
	$G^{JJ}_{,\phi} = \overline{G^{\bar{J}\bar{J}}_{,\phi}}$.
	In the absence of the scalar perturbative sector, the system
	(\ref{SecondOrderEOM1})--(\ref{SecondOrderEOM2}) reduces consistently to the
	dynamics obtained in Ref.~\cite{bojowald2007effective}.
	
	The equations of motion for the mode $v_k$ and its conjugate
	momentum $\pi_k$, extracted from~(\ref{SecondOrderEOM1}), are
	\begin{align}
		v_{k, \phi} &= -\frac{il\gamma p \pi_{k}}{\Delta J} - \frac{i l\gamma p \pi_{k}}{(\Delta J)^{3}}g_{1} + \frac{i l\gamma \pi_{k}}{(\Delta J)^{2}}g_{2}^{-}-\frac{il\gamma}{\Delta J}G^{p\pi} \nonumber \\ &+ \frac{il\gamma p}{(\Delta J)^{2}}g_{4},  \nonumber \\
		\pi_{k, \phi} &= \frac{il\gamma p k^{2}v_{k}}{\Delta J} + \frac{i l\gamma p k^{2}v_{k}}{(\Delta J)^{3}}g_{1} - \frac{i l\gamma k^{2}v_{k}}{(\Delta J)^{2}}g_{2}^{-} + \frac{il\gamma k^{2}}{\Delta J}G^{p v} \nonumber \\ & - \frac{il\gamma k^{2} p}{(\Delta J)^{2}}g_{5},
		\label{eq:mode_eom}
	\end{align}
	where the functions $g_i$ depend only on quantum moments of the
	system and are defined in Appendix~\ref{Appendix}.
	Combining~(\ref{eq:mode_eom}), the mode $v_k$ satisfies the
	second-order equation
	\begin{eqnarray}\label{EffectiveModeDynamics}
		v_{k, \phi \phi} + \mathcal{D} v_{k, \phi}+ \mathcal{W}^{2}v_{k}= 0,  
	\end{eqnarray}
	where $\mathcal{D}$ and $\mathcal{W}^{2}$ are time-dependent coefficients acting as
	effective damping and frequency terms, respectively:
	\begin{eqnarray}\label{DampedFrequencyTerms}
		\mathcal{D} &=& -\frac{A_{, \phi}}{Av_{k, \phi}}(v_{k, \phi}-B) -\frac{AC}{v_{k, \phi}}-\frac{\partial_{\phi}B}{v_{k, \phi}}, \nonumber \\
		\mathcal{W}^{2} &=&k^{2} -U_{\text{eff}},
	\end{eqnarray}
	with $U_{\text{eff}}= -k^{2}(A^{2}-1)$ and  
	\begin{eqnarray}\label{ABC}
		A &:=&  -\frac{il \gamma p}{\Delta J} - \frac{i l\gamma p}{(\Delta J)^{3}}g_{1} + \frac{i l\gamma}{(\Delta J)^{2}}g_{2}^{-}, \nonumber \\
		B &:=&  -\frac{il\gamma}{\Delta J}G^{p\pi} + \frac{il\gamma p}{(\Delta J)^{2}}g_{4}, \nonumber \\
		C &:=&  \frac{il \gamma k^{2}}{\Delta J}G^{p v} - \frac{il\gamma k^{2} p}{(\Delta J)^{2}}g_{5}. 
	\end{eqnarray}
	The damping coefficient $D$ is sourced by the gravitational quantum
	moments $g_1 \sim G^{JJ}$ and $g_2^- \sim G^{J\bar{J}}$
	(see Appendix~\ref{Appendix}), which encode quantum fluctuations
	of the background geometry and act as a dissipative friction on
	the perturbation mode. In the classical limit where all gravitational
	quantum moments vanish, $A\to -ilp/\Delta J\to 1$,
	$B\to 0$, $C\to 0$, and~(\ref{EffectiveModeDynamics}) reduces to the
	Mukhanov--Sasaki equation~(\ref{eq:MS}), providing an important
	consistency check. Far from the bounce, the effective potential
	$U_{\mathrm{eff}}$ is proportional to $p$ and therefore grows with
	the scale factor, leading to an increase in the effective mode
	frequency $\mathcal{W}^{2}$. The quantum moments modify both the oscillation
	amplitude and the frequency of the mode.

	\subsection{Simplified dynamics}
	\label{sec:simplified}
	As a first approximation, we consider the
	test-field case: the scalar mode propagates on a fixed
	semiclassical background, gravitational quantum moments are
	suppressed, and the potential is set to $U = 0$. Setting
	$8\pi G = \gamma = 1$, so that $l^2 = 1/3$, the background
	equations of motion extracted from~(\ref{SecondOrderEOM1}) reduce to
	\begin{align}\label{SimplfiedBackground}
		J_{, \phi}&= \overline{\bar{J}_{, \phi}}= -l  p,  \nonumber \\
		p_{,\phi} &= - \frac{l}{2} \left(J+\bar{J}\right).
	\end{align}
	Although gravitational quantum moment equations are suppressed in
	this truncation, the background variables $p := \langle\hat{p}\rangle$
	and $J := \langle\hat{J}\rangle$ remain expectation values of quantum
	operators. The system therefore does not correspond to a purely
	classical background but rather to a semiclassical one in which
	gravitational fluctuations are suppressed while holonomy corrections are included and quantum effects in
	the matter sector are retained.
	
	The solution of~(\ref{SimplfiedBackground}) is
	\begin{eqnarray}\label{backgroundSolutions}
		p(\phi)	&=& \alpha \cosh(l \phi), \nonumber \\
		J(\phi) &=& -\alpha( \sinh(l\phi)+ i),
	\end{eqnarray}
	where $\alpha$ is a constant proportional to $\hbar$ that sets the
	amplitude of the bounce~\cite{bojowald2007large}. We set $\alpha = 10$
	in natural units ($\hbar = 1$) for the numerical computations of
	Section~\ref{sec:numerical}; the physical observables such as the
	power spectrum and mode frequencies are independent of this choice
	upon appropriate rescaling of units. This solution exhibits a smooth
	bounce since $p = a^2 \geq \alpha > 0$ for all $\phi$, and the
	scale factor evolves as
	\begin{eqnarray}\label{ScaleFactorEvolution}
		a &=& \sqrt{p} = \sqrt{10} \cosh^{1/2}(l \phi).
	\end{eqnarray}
	In a background with $U=0$, cosmic time and the scalar field are
	related by $\phi = \sqrt{2}\,t$.  
	
	In the test-field approximation, the mode equation~(\ref{EffectiveModeDynamics}) reduces to
	\begin{eqnarray}\label{SimplfiedMode}
		v_{k, \phi \phi} +\frac{f_{,\phi}}{f}v_{k,\phi}+ f^{2}k^{2}v_{k} = 0, 
	\end{eqnarray}
	and the quantum moment equations~(\ref{SecondOrderEOM2}) for the scalar
	sector become
	\begin{eqnarray}\label{SecondOrderMomentsEOMsim}
		G^{v v}_{\phi} &=& 2 f G^{v\pi},   \nonumber \\
		G^{\pi \pi}_{\phi}&=&-  2 f k^{2} G^{v\pi} ,  \nonumber \\
		G^{v \pi}_{\phi}&=& f \left(G^{\pi \pi}-k^{2}G^{vv}\right).
	\end{eqnarray}
	where we have defined the dimensionless function
	\begin{equation}
		f \equiv -\frac{il\gamma p}{\Delta J}.
		\label{eq:f_def}
	\end{equation}
	The equality $J_{,\phi} = \overline{\bar{J}_{,\phi}}$ in
	(\ref{SimplfiedBackground}) holds because both quantities share the
	same real part by the symmetry of~(\ref{backgroundSolutions}), while
	their imaginary parts are equal and opposite and therefore contribute
	equally to the derivative.

	Working directly with (\ref{ReducedSolutions}) as deparametrized Hamiltonians is complicated due to the explicit form of the functions given in (\ref{Functionsalphadelta}). Once the assumption of a small amplitude $v_{k}/a \ll 1$ has been made, the Hamiltonian becomes simpler. Nevertheless, the scalar sector behaves as a harmonic oscillator, with the wave number $k$ playing the role of frequency. This approximation is valid at very early stages of inflation, when the modes are deep inside the horizon ($k \gg aH$) and spacetime regions can be considered approximately as regions of Minkowski spacetime \cite{mukhanov2005physical}. However, since the universe is dynamic, its evolution could cause the modes to evolve with a time-dependent frequency \cite{brizuela2019moment}. Under this assumption and from this point on, we relax the condition that $k$ is constant and endow it with a cosmic time dependence, $k^{2} \rightarrow \omega^{2}(t) = (k^{2}-z''/z)$.
	
	\subsubsection{Simplified effective mode $v_{k}$ dynamics}
	\label{sec:mode_dynamics}
	
	Writing~(\ref{SimplfiedMode}) in cosmic time gives
	\begin{eqnarray}\label{tSimplfiedMode}
		\ddot{v}_{k}-\frac{\dot{\mathcal{F}}}{\mathcal{F}}\dot{v}_{k} + \mathcal{F}^{2}\omega^{2}v_{k}=0
	\end{eqnarray}
	where the rescaled function $\mathcal{F}$ is 
	\begin{eqnarray}\label{DimensionlessFfunction}
		\mathcal{F} &:=& -  \sqrt{\frac{2}{3}} \frac{i p }{\Delta J}, 
	\end{eqnarray}
	and the scalar mode evolves with the time-dependent effective
	frequency $\omega^2(t) = k^2 - z''/z$, where $z''/z$ encodes the
	coupling to the background geometry.
	Note that $\mathcal{F}$ and $f$ (defined in Eq.~\eqref{eq:f_def}) are related by
	$F = f/\sqrt{3}$ in the units $8\pi G = \gamma = 1$ used throughout
	this subsection.
	Writing~(\ref{tSimplfiedMode})
	in conformal time gives
	\begin{eqnarray}\label{etaSimplfiedMode}
		v''_{k}-\frac{(a\mathcal{F})'}{a\mathcal{F}}v'_{k} + (a \mathcal{F})^{2}\omega^{2}v_{k}=0.
	\end{eqnarray}
	When $\mathcal{F} = -1/a$, Eq.~(\ref{etaSimplfiedMode}) reduces exactly to
	the standard Mukhanov--Sasaki Eq.~(\ref{eq:MS}). For general
	$\mathcal{F}$, the substitution $v_k = \sqrt{aF}\,u_k$
	transforms~(\ref{etaSimplfiedMode}) into
	\begin{equation}
		u''_{k}+ \left((a\mathcal{F})^{2}k^{2}-\Omega^{2}\right)u_{k} = 0 ,
		\label{ModifiedMukhanovEquation}
	\end{equation}
	with
	\begin{equation}
		\Omega^{2} = (a\mathcal{F})^{2}\frac{z''}{z}+ \frac{\mathcal{Z}''_{\text{eff}}}{\mathcal{Z}_{\text{eff}}},
		\qquad
		\mathcal{Z}_{\text{eff}} = (a\mathcal{F})^{-1/2}.
		\label{eq:Omega}
	\end{equation}
	
	Equation~(\ref{ModifiedMukhanovEquation}) preserves the Mukhanov--Sasaki
	structure~(\ref{eq:MS}) but with modified background functions
	$(a \mathcal{F})$ and $\omega^2$, which encode the LQC bounce corrections
	through the function $\mathcal{F}$.
	\subsubsection{Simplified quantum moments dynamics}
	Combining the three equations in~(\ref{SecondOrderMomentsEOMsim}), the
	mean squared dispersion $G^{vv}$ satisfies the third-order ordinary
	differential equation
	\begin{eqnarray}\label{ThirdOrderEq1}
		\dddot{G}^{vv}-3\frac{\dot{\mathcal{F}}}{\mathcal{F}} \ddot{G}^{vv} + \left(4 \mathcal{F}^{2}\omega^{2} -\frac{\ddot{\mathcal{F}}}{\mathcal{F}}+3\left(\frac{\dot{\mathcal{F}}}{\mathcal{F}}\right)^{2} \right)\dot{G}^{vv}\nonumber \\
		+4 \mathcal{F}^{2}\omega\dot{\omega}G^{vv} = 0.
	\end{eqnarray}
	Introducing the conformal variable $\xi = -k\eta$ and
	writing~(\ref{ThirdOrderEq1}) in conformal time yields
	\begin{widetext}
		\begin{eqnarray}\label{ThirdOrderEq2}
			k^{3}G^{vv}_{,\xi \xi \xi}-3k^{3}\frac{(a\mathcal{F})_{,\xi}}{a\mathcal{F}} G^{vv}_{,\xi \xi} + \left[4 k a^{2} \mathcal{F}^{2}\omega^{2} +k^{3}\left(3\left(\frac{(a\mathcal{F})_{,\xi}}{a\mathcal{F}} \right)^{2} - 2\frac{a_{,\xi}\mathcal{F}_{,\xi}}{a\mathcal{F}}  - \frac{a_{,\xi \xi}}{a}-\frac{\mathcal{F}_{,\xi \xi}}{\mathcal{F}} \right)\right]G^{vv}_{,\xi} +4 k a^{2}\mathcal{F}^{2}\omega \omega_{,\xi} G^{vv}  = 0 .\nonumber \\
		\end{eqnarray}
	\end{widetext}	
	Setting again $\mathcal{F}=-1/a$ in~(\ref{ThirdOrderEq2}), the second term
	and the parenthetical expression in the bracket both vanish
	identically, and the equation reduces to
	\begin{eqnarray}
		k^{3}G^{vv}_{,\xi \xi \xi}+ 4 k \omega^{2}G^{vv}_{,\xi} +4 k \omega \omega_{,\xi} G^{vv}  = 0 .
		\label{eq:Gvv_singular}
	\end{eqnarray}
	Substituting the de~Sitter frequency
	$\omega^2 = k^2(1 - 2/\xi^2)$ into~(\ref{eq:Gvv_singular}) gives
	\begin{eqnarray}\label{NoHolonomyCorrectionEquation}
		\xi^{3} G^{vv}_{,\xi \xi \xi}+ 4 \xi \left(\xi^{2}-2\right) G^{vv}_{,\xi} + 8 G^{vv}  = 0,
	\end{eqnarray}
	which is precisely the equation derived in Ref.~\cite{brizuela2019moment}
	for the evolution of $G^{vv}$ in a singular classical de~Sitter
	background. Equations~(\ref{etaSimplfiedMode})
	and~(\ref{ThirdOrderEq2}) therefore generalize the results of
	Ref.~\cite{brizuela2019moment} to the bouncing background considered here.
	
	\subsubsection{Classical limit: $\mathcal{F}=-1/a$} \label{subsec:classicallimit}
	The physical meaning of the condition $\mathcal{F}=-1/a$ becomes clear when
	the $\mu$-dependence suppressed in Section~\ref{sec:LQC} is
	restored. Writing $\Delta J = J - \bar{J} = (2ip/\mu)\sin(\mu c)$
	and substituting into~(\ref{DimensionlessFfunction}), we obtain
	\begin{eqnarray}
		\mathcal{F} &=& -  \sqrt{\frac{2}{3}} \frac{i p }{\Delta J} = -  \sqrt{\frac{2}{3}} \frac{\mu}{2 \sin(\mu c)}.  
	\end{eqnarray}
	In the low-curvature limit $\mu c \to 0$ (equivalently, $|c|\ll \mu^{-1}$), $\sin(\mu c) \sim \mu c$,
	\begin{equation}\label{1/c}
		\mathcal{F} =- \frac{1}{\sqrt{6}} \frac{1}{c} .  
	\end{equation}
	In the de~Sitter regime, the scale factor~(\ref{ScaleFactorEvolution})
	behaves as
	\begin{eqnarray}\label{scaleFactordeSitter}
		a \approx \sqrt{5} \exp(\frac{\phi}{2\sqrt{3}}) = \sqrt{5}\exp(\frac{t}{\sqrt{6}}),
	\end{eqnarray}
	so its time derivative is $c = \dot{a} = a/\sqrt{6}$, and
	(\ref{1/c}) gives $\mathcal{F} = -1/a$. Therefore,
	Eqs.~(\ref{etaSimplfiedMode}) and~(\ref{ThirdOrderEq2}) reduce
	to~(\ref{eq:MS}) and~(\ref{NoHolonomyCorrectionEquation}), respectively, in the
	low-curvature, late-time regime where $\mu\to 0$ and the bounce
	contribution is negligible. The function $\mathcal{F}$ encodes both the
	semiclassical bounce correction (through $\mu$ and $\sin(\mu c)$)
	and the classical de~Sitter limit ($\mathcal{F} = -1/a$) within a single
	unified expression.

	\subsection{Quantum scalar sector in a free classical de Sitter spacetime}
	\label{sec:dS_bounce}
	We now compute the correction to the dimensionless power spectrum
	arising from the LQC bounce, extending the calculation of
	Ref.~\cite{brizuela2019moment} from a singular to a bouncing background. 
	For large $|\phi|$, the system enters the de~Sitter regime and
	the scale factor grows exponentially. The de~Sitter behavior 
	is an asymptotic property of the
	bounce solution~\eqref{backgroundSolutions} that holds independently of the value of the
	potential; setting $U=0$ does not affect this regime.
	Using~(\ref{scaleFactordeSitter}), the
	conformal-time form of the scale factor is
	\begin{eqnarray}
		a \approx \sqrt{5}\exp(\frac{t}{\sqrt{6}}) = -\frac{\sqrt{6}}{\eta},
	\end{eqnarray}
	and the function $\mathcal{F}$ from~(\ref{DimensionlessFfunction}), evaluated on the
	background solution~(\ref{SimplfiedBackground}), becomes
	\begin{equation}
		\mathcal{F} = \frac{1}{\sqrt{6}}\cosh(l\phi)
		\approx \frac{3}{5\sqrt{6}\eta^2}
		= \frac{3k^2}{5\sqrt{6}\xi^2},
		\label{eq:F_dS}
	\end{equation}
	where in the last step we used $\eta = -1/(k\xi)$. The effective
	frequency $\omega$ takes the standard de~Sitter form
	\begin{equation}
		\omega^{2} \approx k^{2}-\frac{2}{\eta^{2}}.
		\label{eq:omega_dS}
	\end{equation}
	Collecting these expressions, the de~Sitter approximations in terms
	of $\xi = -k\eta$ are
	\begin{align}\label{DeSitterFunctions}
		a& \approx\sqrt{6}\frac{k}{\xi},	\nonumber \\
		\mathcal{F}&\approx	\frac{3 k^{2}}{5\sqrt{6}\xi^{2}},	\nonumber \\
		\omega^{2}&\approx k^{2}\left(1-\frac{2}{\xi^{2}}\right).
	\end{align}
	Substituting~(\ref{DeSitterFunctions}) into~(\ref{ThirdOrderEq2}),
	the equation for $G^{vv}$ in the de~Sitter bouncing background
	becomes
	\begin{eqnarray}\label{ThirdOrderEq3}
		\xi^{3} \frac{d^{3}G^{vv}}{d\xi^{3}}+9\xi^{2} \frac{d^{2}G^{vv}}{d\xi^{2}} + \left(15\xi +\frac{36 k^{6}(\xi^{2}-2)}{25 \xi^{5}}\right)\frac{dG^{vv}}{d\xi}  \nonumber \\
		+\frac{72 k^{6}}{25 \xi^{6}} G^{vv}  = 0. \nonumber \\
	\end{eqnarray}
	Adding and subtracting the terms of the singular-background
	equation~(\ref{NoHolonomyCorrectionEquation}), this can be rewritten as
	\begin{align}\label{ThirdOrderEq4}
		&\xi^{3} \frac{d^{3} G^{vv}}{d\xi^{3}} + 4\xi (\xi^{2}-2)\frac{d G^{vv}}{d\xi} +8 G^{vv} \nonumber \\
		&+9\xi^{2} \frac{d^{2}G^{vv}}{d\xi^{2}} + \left(15\xi +\frac{36 k^{6}(\xi^{2}-2)}{25 \xi^{5}} - 4\xi \left(\xi^{2}-2\right)\right)\frac{dG^{vv}}{d\xi}  \nonumber \\
		&+8\left(\frac{9 k^{6}}{25 \xi^{6}}-1\right) G^{vv}  = 0. \nonumber \\
	\end{align}
	The first line of~(\ref{ThirdOrderEq4}) is operator
	$\mathcal{A}$ of Ref.~\cite{brizuela2019moment}. The remaining
	terms define the operator $\mathcal{B}$, which encodes the LQC bounce
	correction. Equation~(\ref{ThirdOrderEq4}) therefore generalizes the
	result of Ref.~\cite{brizuela2019moment} to the bouncing background.
	
	Assuming a Bunch--Davies vacuum as the initial condition (physically
	motivated because the modes are deep inside the Hubble horizon
	at early times, where spacetime is approximately flat), the solution
	to the singular-background equation
	\begin{eqnarray}\label{Brizuela'sThirdOrderEquation}
		\xi^{3} \frac{d^{3} G^{vv}_{\text{s}}}{d\xi^{3}} + 4\xi (\xi^{2}-2)\frac{d G^{vv}_{\text{s}}}{d\xi} +8 G^{vv}_{\text{s}}&=&0,
	\end{eqnarray}
	where the subscript $s$ stands for the singular background, is   
	\begin{eqnarray}\label{KnownSolution}
		G^{vv}_{\text{s}}&=& \frac{\hbar}{2k} \left(\frac{1+\xi^{2}}{\xi^{2}}\right).
	\end{eqnarray}
	
	We seek a solution of~(\ref{ThirdOrderEq3}) assuming that the cosmic bounce produces a small correction $\mathcal{G}^{vv}$ to $G^{vv}_{\text{s}}$, i.e., a solution of the form
	\begin{eqnarray}\label{CorrectedSolution}
		G^{vv}_{\text{Bounce}} &=& G^{vv} + \lambda \mathcal{G}^{vv},
	\end{eqnarray}
	where $\lambda$ is a perturbation parameter that tracks the
	order of the LQC correction; setting $\lambda = 1$ at the end
	recovers the physical result. Substituting~(\ref{CorrectedSolution})
	into~(\ref{ThirdOrderEq3}) and using the operator decomposition
	$(\mathcal{A} + \lambda\mathcal{B})(G^{vv}_s + \lambda\widetilde{G}^{vv}) = 0$,
	we obtain
	\begin{eqnarray}\label{PerturbativeEquation1}
		\mathcal{A}(G^{vv}_{\text{s}}) + \mathcal{A}(\mathcal{G}^{vv})  + \lambda \mathcal{B}(G^{vv}_{\text{s}}) + \lambda^{2}\mathcal{B}(\mathcal{G}^{vv}) &=&0.
	\end{eqnarray}
	The first term vanishes by construction. Retaining terms to first
	order in $\lambda$ yields the inhomogeneous equation
	\begin{equation}
		\mathcal{A}(\mathcal{G}^{vv}) = -\lambda\mathcal{B}(G^{vv}_s).
		\label{PerturbativeEquation2}
	\end{equation}
	Evaluating $\mathcal{B}$ on the known
	solution~(\ref{KnownSolution}), the right-hand side of
	(\ref{PerturbativeEquation2}) gives
	\begin{equation}\label{ThirdOrderEq5}
		\xi^{3} \frac{d^{3} \mathcal{G}^{vv}_{\text{s}}}{d\xi^{3}} + 4\xi (\xi^{2}-2)\frac{d \mathcal{G}^{vv}_{\text{s}}}{d\xi} +8 \mathcal{G}^{vv}_{\text{s}} =- \frac{108 \hbar k^{5}}{25 \xi^{8}}.
	\end{equation}
	Imposing Bunch--Davies initial conditions, which select the unique
	solution that reduces to the standard vacuum fluctuation in the
	sub-horizon limit $\xi\to\infty$ and fixing the three integration
	constants of~(\ref{ThirdOrderEq5}), the general solution is
	\begin{widetext}
		\begin{eqnarray}
			\mathcal{G}^{vv} &=& \left[ \frac{\hbar}{2k}\left(\frac{1+\xi^{2}}{\xi^{2}}\right)- \frac{16 \hbar k^{5}\pi}{78750}\left(\frac{\sin(\xi)}{\xi} -\cos(\xi)\right)\left(\frac{\cos(\xi)}{\xi}+\sin(\xi)\right)\right]
			+ \frac{\hbar k^{5}}{78750} \left\lbrace \frac{525}{\xi^{8}}-\frac{60}{\xi^{6}}+\frac{18}{\xi^{4}}-\frac{20}{\xi^{2}}  +\right.\nonumber \\
			&&\left. 16\left[\left(\frac{1-\xi^{2}}{\xi^{2}}\right)\left(\cos(2\xi)C_{\text{i}}[2\xi]+\sin(2\xi)S_{\text{i}}[2\xi]\right)+\frac{2}{\xi^{2}}\left(\sin(2\xi)C_{\text{i}}[2\xi]-\cos(2\xi)S_{\text{i}}[2\xi]\right)	\right]\right \rbrace,
		\end{eqnarray}
	\end{widetext}
	where $\mathrm{Ci}$ and $\mathrm{Si}$ are the cosine and sine
	integral functions, respectively.
	
	We evaluate~(\ref{CorrectedSolution}) in the late-time limit
	$a\to\infty$ (equivalently $\xi\to 0$). The cosine integral
	$\mathrm{Ci}(2\xi)$ diverges logarithmically as $\xi\to 0$,
	which signals that the perturbative correction $\mathcal{G}^{vv}$
	does not admit a smooth super-horizon freeze-out in the same way as
	the classical mode function. This is a limitation of the
	perturbative treatment: the bounce correction modifies the
	super-horizon dynamics of $G^{vv}$, and a complete description of
	this regime would require either resummation of the perturbative
	series or a fully numerical treatment. As a representative estimate
	of the correction at the moment of horizon crossing, we therefore
	evaluate~(\ref{CorrectedSolution}) at $\xi = 1$, corresponding to
	$k|\eta| = 1$. In terms of the scale factor, the result is
	\begin{align}\label{SolutionIncludingBounce}
		&G^{vv}_{\text{Bounce}}|_{a\rightarrow \infty} \approx  \frac{\hbar a^{2}H^{2}}{k^{3}}
		+ \frac{\hbar }{78750} \left[ \frac{525}{k^{3}}(aH)^{8}\right.\nonumber \\
		& \left.  -\frac{60}{k}(aH)^{6}+18(aH)^{4}k  + 4\left(4C_{\text{i}}[2]-5\right)(aH)^{2}k^{3}\right].
	\end{align}
	Using~(\ref{SolutionIncludingBounce}), the bounce-corrected dimensionless
	curvature power spectrum~\cite{brizuela2019moment} 
	\begin{eqnarray}\label{PowerspectrumBounce}
		\mathcal{P}_{\text{Bounce}}&\approx&\frac{Gk^{3}}{a^{2}}G_{\text{Bounce}}^{vv}.
	\end{eqnarray}
	is obtained by reintroducing the fundamental constants $G$ and
	$\hbar$.
	\begin{align}\label{PowerSpectrumCorrection}
		\mathcal{P}_{\text{Bounce}} &= \tilde{H}^{2}\left\lbrace1 
		+ \frac{\ell_{\mathrm{Pl}}^{6} }{78750} \left[ 525(aH)^{6}-60(aH)^{4}k^{2} \right. \right. \nonumber \\
		& \left. \left. +18(aH)^{2}k^{4}  + 4\left(4C_{\text{i}}[2]-5\right)k^{6}\right] \right\rbrace,
	\end{align}
	where $\tilde{H} = \ell_{\mathrm{Pl}}^2 H^2  =(H/H_{\rm Pl})^2$ is the dimensionless
	Hubble parameter and $\ell_{\mathrm{Pl}} = \sqrt{G\hbar/c^3}$ is
	the Planck length.
	
	Several features of~(\ref{PowerSpectrumCorrection}) deserve comment. First,
	the correction is suppressed by $\ell_{\mathrm{Pl}}^6$, confirming
	that LQC effects are rapidly diluted by cosmic expansion and are
	negligible at all cosmologically observable scales. Second, the
	correction is scale-dependent: at horizon crossing $aH = k$, all
	terms reduce to the same order $k^6\ell_{\mathrm{Pl}}^6$, so the
	corrected spectrum takes the schematic form
	$\mathcal{P}_\mathcal{R}(k) \approx \tilde{H}^2
	[1 + C(k\ell_{\mathrm{Pl}})^6]$, where $C$ is a
	numerical coefficient. The spectral tilt receives a correction
	\begin{equation}
		\delta n_s =
		\frac{d\ln\mathcal{P}_{\mathrm{Bounce}}}{d\ln k}
		\approx 6C(k\ell_{\mathrm{Pl}})^6 \ll 1,
		\qquad\text{for }k\ell_{\mathrm{Pl}} \ll 1,
		\label{eq:tilt}
	\end{equation}
	which is entirely negligible for all observable modes, consistent
	with the Planck measurement $n_s = 0.965 \pm 0.004$. Here $C$ collects the contributions of all four spectral terms in~(75)
	evaluated at horizon crossing $aH=k$; an explicit computation gives
	\begin{equation*}
		C = \frac{1}{78750}\bigl[525 - 60 + 18 +
		4(4\,{\rm Ci}[2]-5)\bigr]\approx 6.1\times10^{-3},
	\end{equation*}
	yielding $\delta n_s\lesssim 4\times10^{-20}$ at the CMB pivot scale
	$k_*=0.05\,{\rm Mpc}^{-1}$.
	
	Third, the correction obtained in this analysis, proportional to $k^6\ell_{\mathrm{Pl}}^6$, corresponds to the inclusion of bounce effects within the test-field approximation, where quantum perturbations evolve on an effective de Sitter background modified by the bounce, without backreaction.
	
	Fourth, the
	absence of the discretization parameter $\mu$ in~(\ref{PowerSpectrumCorrection})
	is a consequence of the de~Sitter approximation: in this regime
	$\mathcal{F}$ is evaluated at late times where the holonomy correction
	$\sin(\mu c)/\mu\to c$ and $\mu$ drops out. In more general
	scenarios, such as the slow-roll approximation, $\mu$ would
	appear explicitly. Its absence here reflects that the bounce
	information is encoded in the initial conditions of the de~Sitter
	phase rather than in the functional form of $\mathcal{F}$ at late times.
	
	It should be emphasized that the
	enhancement in~(\ref{PowerSpectrumCorrection}) is computed within the
	test-field approximation, in which the gravitational quantum
	moments are suppressed. The numerical analysis of
	Section~\ref{sec:numerical} shows that when the full second-order
	dynamics is considered, gravitational quantum fluctuations introduce
	an effective friction through the damping term $\mathcal{D}$ in
	Eq.~(\ref{EffectiveModeDynamics}), which suppresses the scalar perturbation
	amplitude after the bounce. These two results are complementary: Eq.~(\ref{PowerSpectrumCorrection}) captures the
	kinematic imprint of the bounce on the power spectrum within the
	test-field approximation and the numerical results reveal the
	dynamical suppression due to quantum backreaction when gravitational
	moments are retained.

	\section{Numerical evolution}\label{sec:numerical}
	
	The dynamics given by~(\ref{SecondOrderEOM1}) and (\ref{SecondOrderEOM2})
	constitutes a system of twenty coupled, highly nonlinear ordinary
	differential equations describing the evolution of a scalar
	perturbation mode propagating on a nonsingular gravitational
	background, together with the second-order quantum moments of the
	system. The strong coupling between the gravitational and matter
	sectors, combined with the technical complexity of the equations of
	motion, makes an analytical closed-form solution unavailable. We
	therefore solve the system numerically, setting the scalar potential
	$U = 0$ throughout this section. This choice is physically motivated:
	near the cosmic bounce the potential is subdominant relative to the
	kinetic energy and can be neglected at the classical level.
	
	It is useful to organize the analysis according to a hierarchy of
	approximations to the full second-order effective dynamics.
	\textit{Level~1} corresponds to the test-field approximation of
	Section~\ref{sec:simplified}, in which gravitational quantum moments are
	suppressed and the scalar mode propagates on a fixed semiclassical
	background; this level yields the analytical
	correction~(\ref{PowerSpectrumCorrection}).
	\textit{Level~2} retains the dynamical evolution of the gravitational
	quantum moments through equations $G_{1}$, $G_{2}$,
	$G_{5}$, $G_{6}$ and their effect on the mode
	equations $F_{v_k}$, $F_{\pi_k}$, but the cross-sector moments
	$G^{Jv}$, $G^{J\pi}$, $G^{pv}$, $G^{p\pi}$ are kept identically
	zero throughout the evolution; this defines a quantum extension of
	the test-field approximation in which the geometry's quantum
	fluctuations affect the mode, but the mode does not generate quantum
	correlations with the geometry.
	\textit{Level~3} is the full second-order system, in which the
	cross-sector moments are also evolved through
	$G_{7}$--$G_{10}$.
	The main numerical results of this section correspond to Level~2;
	the comparison with Level~3 is presented at the end of
	Section~\ref{sec:numerical}.
	
	\subsection{Initial conditions}
	\label{sec:ICs}
	
	\subsubsection{Gravitational sector}
	The gravitational background variables must satisfy initial
	conditions that guarantee the existence of a cosmic bounce with a
	finite, nonvanishing minimum value of the scale factor. Since
	$p = a^2$, consistency with the background solution~(\ref{ScaleFactorEvolution})
	requires $p_0 = \alpha = 10$. This choice partially fixes the
	non-canonical variables $J = pe^{ic}$ and $\bar{J} = pe^{-ic}$,
	leaving only the initial phase $c_0$ to be specified. We set
	$c_0 = \pi/4$, which ensures that the combinations $J + \bar{J}$
	(purely real) and $J - \bar{J} = \Delta J$ (purely imaginary) are
	both nonvanishing at $\phi = 0$. This is essential because many
	terms in~(\ref{SecondOrderEOM1}) and~(\ref{SecondOrderEOM2}) contain inverse
	powers of $\Delta J$, which would otherwise be ill-defined. The
	gravitational initial conditions are therefore
	\begin{equation}\label{poJoIC}
		p_{0}= 10 , \hspace{0.4cm}  J_{0}=\frac{10}{\sqrt{2}}\left (1+ i\right)  , \hspace{0.4cm}  \bar{J}_{0}= \frac{10}{\sqrt{2}} \left(1-i\right).
	\end{equation}
	\subsubsection{Matter sector}
	For the scalar mode we impose the Bunch--Davies vacuum as initial
	condition. This choice has a physical motivation: at very early
	times, close to the bounce and at the onset of inflation, the
	perturbation modes are deep inside the Hubble horizon ($k \gg aH$)
	and spacetime is approximately flat. The standard Bunch--Davies
	initial conditions are~\cite{baumann2009tasi}
	\begin{equation}\label{BunchDavies}
		v_{k,0}= \frac{1}{\sqrt{2k}} e^{-ik \eta_{0}}, \hspace{0.4cm} \pi_{k, 0}= -i \sqrt{\frac{k}{2}} e^{-ik \eta_{0}}.
	\end{equation}
	\subsubsection{Quantum moments}
	For the second-order quantum moments we employ the Gaussian
	state~(\ref{GaussP}), which is a minimum-uncertainty state
	and represents the closest quantum analogue to a classical
	configuration while still encoding nontrivial quantum fluctuations.
	More general, non-Gaussian initial states are in principle
	possible; however, the Gaussian choice is physically motivated
	at the bounce scale and provides a natural starting point for the
	analysis. Up to second order, the initial moments for the canonical
	variables $c$ and $p$ are
	\begin{eqnarray}\label{originalQMcp}
		G^{cc}_{0}= \sigma^{2}, \hspace{1cm} G^{cp}_{0}=0, \hspace{1cm} G^{pp}_{0}= \frac{1}{4\sigma^{2}}.
	\end{eqnarray}
	where $\sigma$ is the width of the Gaussian wave packet for the
	gravitational sector.
	
	The initial conditions for the moments of the non-canonical
	variables $J$ and $\bar{J}$ are obtained by expanding $\hat{J}$ to
	first order in quantum fluctuations around its expectation value:
	\begin{eqnarray}
		\hat{J}&=&J(\hat{p},\hat{c})= J(\Delta{p} + \langle \hat{p} \rangle,\Delta{c} + \langle \hat{c}\rangle) \nonumber \\
		&\approx& \langle \hat{J} \rangle +J_{,p}\Delta{p} +J_{,c}\Delta{c} \nonumber \\ 
		&=& \langle \hat{J} \rangle +\frac{J}{p}\Delta{p} +iJ\Delta{c},
		\label{eq:J_expand}
	\end{eqnarray}
	where $\Delta p = \hat{p} - \langle\hat{p}\rangle$ and
	$\Delta c = \hat{c} - \langle\hat{c}\rangle$, and we used
	$J_{,p} = J/p$ and $J_{,c} = iJ$. Squaring~(\ref{eq:J_expand})
	and taking the expectation value gives
	\begin{equation}
		G^{JJ}_0
		= \langle(\hat{J} - \langle\hat{J}\rangle)^2\rangle_0
		= J_0^2\!\left(
		\frac{G^{pp}_0}{p_0^2}
		+ \frac{2i}{p_0}G^{cp}_0
		- G^{cc}_0
		\right).
		\label{eq:GJJ_from_canonical}
	\end{equation}
	Substituting~(\ref{poJoIC}) and~(\ref{originalQMcp}),
	\begin{equation}
		G^{JJ}_0 =
		\left(\frac{10}{\sqrt{2}}(1+i)\right)^{\!2}
		\left(\frac{1}{4(10\sigma)^2} - \sigma^2\right).
		\label{eq:GJJ_explicit}
	\end{equation}
	Note that since $(1+i)^2 = 2i$, $G^{JJ}_0$ is purely imaginary.
	This is not unexpected: $J = pe^{ic}$ is not self-adjoint, so
	its second moment $G^{JJ} = \langle(\hat{J}-\langle\hat{J}\rangle)^2\rangle$
	is not required to be real. The physically relevant dispersions
	are the real quantities $G^{J\bar{J}}$ and $G^{JJ}$, which
	appear in the equations of motion. Proceeding analogously for
	the remaining non-canonical moments, the complete set of initial
	gravitational quantum moments is
	\begin{align}
		G^{JJ}_0 = \overline{G^{\bar{J}\bar{J}}_0}
		&= \left(\frac{10}{\sqrt{2}}(1+i)\right)^{\!2}
		\!\left(\frac{1}{4(10\sigma)^2} - \sigma^2\right),
		\notag\\
		G^{Jp}_0 = \overline{G^{\bar{J}p}_0}
		&= \frac{1}{4\sqrt{2}\,\sigma^2}(1+i),
		\notag\\
		G^{J\bar{J}}_0
		&= \frac{1}{4\sigma^2} + (10\sigma)^2,
		\notag\\
		G^{pp}_0
		&= \frac{1}{4\sigma^2}.
		\label{InicialConditons1}
	\end{align}
	For the scalar perturbation mode, the Gaussian initial state with
	width $\chi$ gives
	\begin{equation}\label{InicialConditons2}
		G^{vv}_{0}= \chi^{2}, \hspace{1cm} G^{v\pi}_{0}=0, \hspace{1cm} G^{\pi \pi}_{0}= \frac{1}{4\chi^{2}},
	\end{equation}
	$\chi$ is the dispersion of the wave packet associated with
	the scalar mode and is independent of the gravitational dispersion
	$\sigma$. As shown in Ref.~\cite{hernandez2024singularity} and
	established in the context of quantum backreaction
	in~\cite{hernandez2026quantum}, nontrivial and physically
	relevant results can be obtained even in the absence of initial
	gravitational-matter quantum correlations. We therefore set all
	cross-sector moments to zero initially,
	\begin{equation}\label{InicialConditons3}
		G^{J v}_{0} =G^{\bar{J}v}_{0} = G^{J\pi}_{0} =G^{\bar{J} \pi}_{0} = G^{p v}_{0} = G^{p \pi}
		= 0 .
	\end{equation}
	These cross-moments will generically become nonzero during the
	evolution, since the equations $G_7$--$G_{10}$
	in~(\ref{SecondOrderEOM2}) couple them to $G^{JJ}$, $G^{pp}$, $G^{vv}$,
	and $G^{v\pi}$. Their dynamical generation encodes the quantum
	backreaction between the bounce geometry and the scalar mode, and
	represents an important effect for future
	investigation~\cite{hernandez2026quantum}. The full numerical
	evolution is obtained using~(\ref{InicialConditons1}),
	(\ref{InicialConditons2}), and~(\ref{InicialConditons3}) as initial
	conditions.
	\subsection{Numerical evolution}
\begin{figure}[H]
		\centering
		\includegraphics[width=0.45\textwidth]{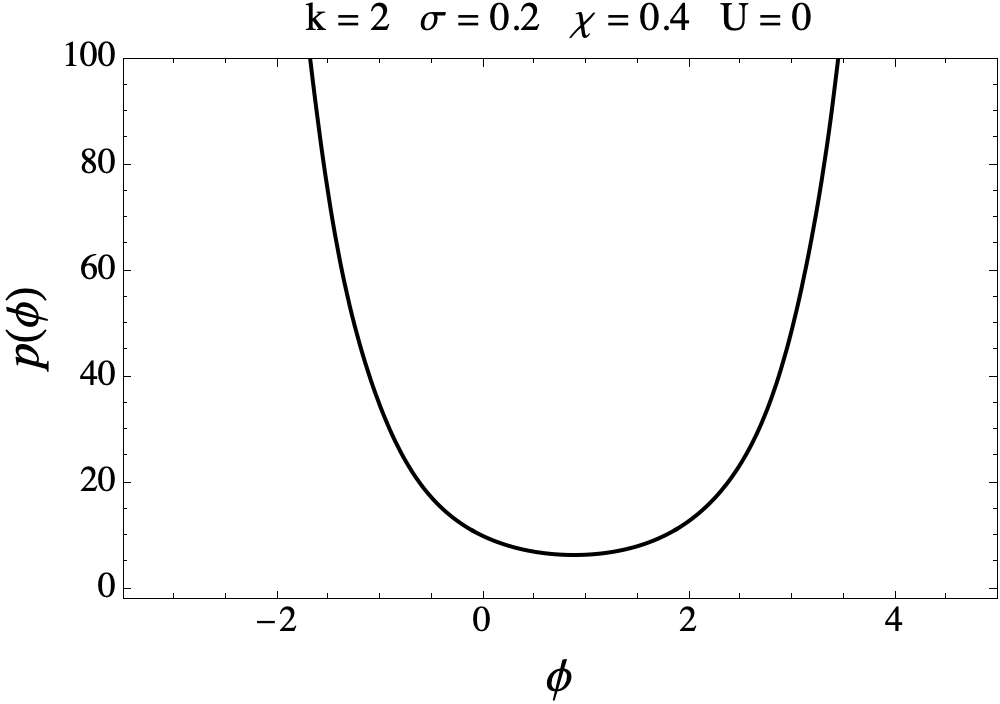}
		\caption{Evolution of $p(\phi) = a^2(\phi)$ as a function of the internal time $\phi$ (scalar field) for $k=2$, $\sigma=0.2$,
			$\chi=0.4$, $U=0$. The plot exhibits a smooth bounce at $\phi=0$,
			where $p$ reaches its minimum value $p_{\min}=\alpha=10$. For
			$|\phi|\gg 2$ the scale factor grows exponentially, approaching
			the de~Sitter regime described by Eq.~(\ref{ScaleFactorEvolution}).}
		\label{Bounce}
	\end{figure}

\begin{figure}[H]
		\centering
		\includegraphics[width=0.45\textwidth]{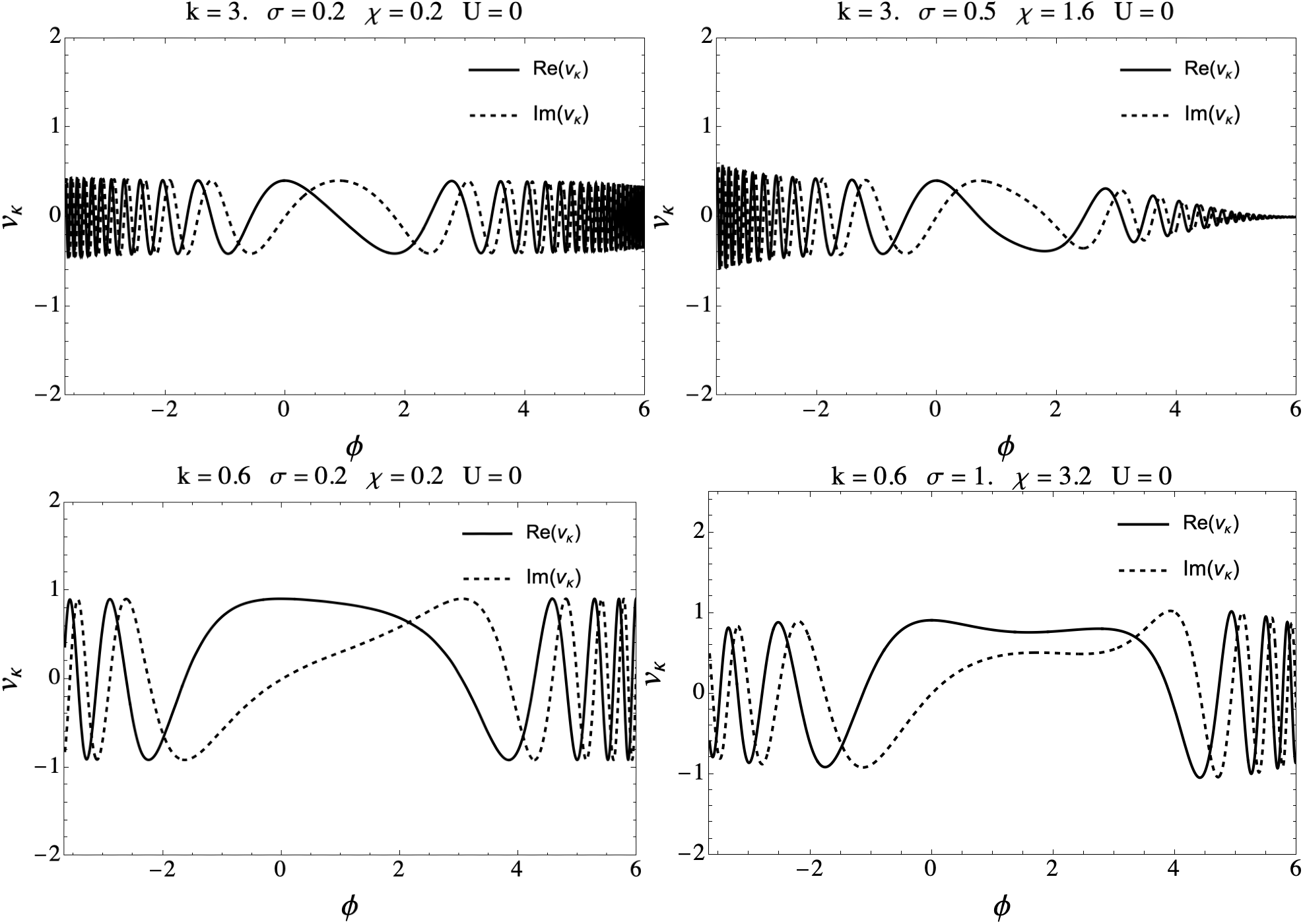}
		\caption{Evolution of $\mathrm{Re}(v_k)$ and $\mathrm{Im}(v_k)$
			through the bounce ($0\lesssim\phi\lesssim 2$) for modes
			$k=3$ (top row) and $k=0.6$ (bottom row), at fixed $\chi$ and
			two values of the gravitational dispersion $\sigma$. The
			horizontal axis shows the internal time $\phi$ (scalar field), with
			$\phi=0$ marking the bounce. During contraction ($\phi\lesssim 0$)
			the mode oscillates with increasing frequency as $|\phi|$
			grows. The amplitude reaches a minimum at the bounce and resumes
			growth during the expansion phase ($\phi\gtrsim 2$). Increasing
			$\sigma$ enhances the damping of the oscillation amplitude after
			the bounce, consistent with the interpretation of gravitational
			quantum moments as an effective friction mechanism.}
		\label{ModesEvolution}
	\end{figure}

	Figure~\ref{Bounce} shows the evolution of $p = a^2$ as a
	function of the internal time $\phi$ (the scalar field), which
	serves as the clock in the deparametrized system. The plot exhibits
	a smooth bounce occurring in the region $0 \lesssim \phi \lesssim 2$,
	where the scale factor reaches its minimum value $a_{\min} =
	\sqrt{\alpha} = \sqrt{10}$ at $\phi = 0$. For $|\phi| \gg 2$
	the scale factor grows exponentially, consistent with the de~Sitter
	behavior described by~(\ref{ScaleFactorEvolution}), and in agreement
	with the analytical background solution~(\ref{backgroundSolutions}).
	Figure~\ref{ModesEvolution} shows the real and imaginary parts of two
	scalar modes, $k=0.6$ and $k=3$, passing through the bounce for
	different values of the gravitational dispersion $\sigma$ at fixed
	scalar dispersion $\chi$. The evolution naturally separates into
	three phases: contraction, bounce, and expansion. 
	\subsubsection{Contraction phase: $\phi \lesssim 0$}
	During contraction the mode oscillates with effective frequency
	$\mathcal{W}^{2} = k^{2}-U_{\text{eff}}$ given by~(\ref{DampedFrequencyTerms}), with $k^2$
	dominating over $U_{\mathrm{eff}}$ for sub-horizon modes. The
	coupling between the mode and the scale factor results in a
	progressively increasing oscillation frequency as $\phi$ becomes
	more negative. Modes with larger wavenumber $k$ oscillate more
	rapidly, as expected. The gravitational quantum moments, which grow
	with $\sigma$, appear in the dynamics as friction terms through the
	damping coefficient $\mathcal{D}$ in~(\ref{EffectiveModeDynamics}) and produce a slight
	modification of the mode amplitude. For larger modes ($k=3$),
	quantum gravitational effects enhance the amplitude before the
	bounce; for smaller modes ($k=0.6$), the amplitude is slightly
	reduced.
	
	\subsubsection{Bounce phase: $0 \lesssim \phi \lesssim 2$}
	As the Universe contracts toward the bounce, the scale factor
	reaches its minimum at $\phi \approx 1$ and the effective frequency
	$\mathcal{W}^2$ is significantly reduced. The mode oscillation therefore
	slows and the perturbation amplitude reaches a local minimum when
	crossing the bounce. Once the bounce occurs and expansion begins,
	the modes resume oscillating. This is qualitatively different from the singular case: without a bounce, the scale factor collapses to zero, the mode frequency
	vanishes, and the mode freezes permanently at the singularity.
	
	\subsubsection{Expansion phase: $\phi \gtrsim 2$}
	During the expansion phase, the interplay between the bounce
	dynamics and the gravitational quantum moments determines the
	evolution of the perturbation amplitude.
	The effective friction coefficient~$\mathcal{D}$
	in~(\ref{DampedFrequencyTerms}), sourced by the gravitational
	quantum moments $g_1 \sim G^{JJ}$ and $g_2^- \sim G^{J\bar{J}}$,
	competes with the bounce-induced amplification that acts on modes
	whose physical wavelength is comparable to the bounce scale.
	
\begin{figure}[H]
		\centering
		\includegraphics[width=0.43\textwidth]{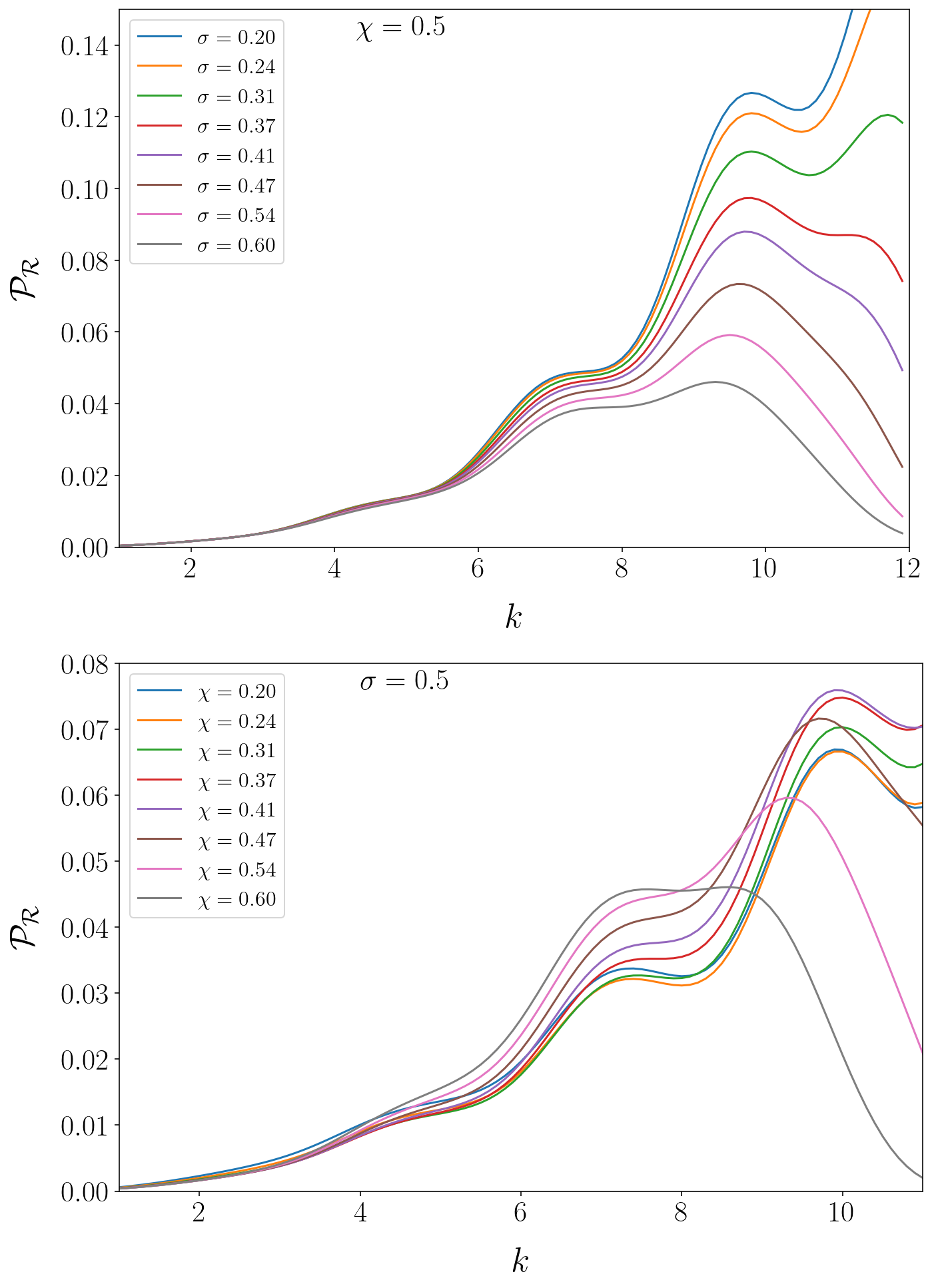}
		\caption{Power spectrum $\mathcal{P}_\mathcal{R}(k)$ at
			$\phi = 2$ within Level~2, with $U = 0$.
			\textit{Upper panel}: dependence on the gravitational
			dispersion~$\sigma$ at fixed $\chi = 0.5$. Curves correspond
			to $\sigma \in \{0.20, 0.24, 0.31, 0.37, 0.41, 0.47, 0.54,
			0.60\}$. The transition between the weak-dispersion regime
			(monotonic ultraviolet growth) and the strong-dispersion regime
			(peak followed by suppression) occurs at
			$\sigma_c \approx 0.24$--$0.37$.
			\textit{Lower panel}: dependence on the scalar dispersion~$\chi$
			at fixed $\sigma = 0.5$. Curves correspond to $\chi \in
			\{0.20, 0.24, 0.31, 0.37, 0.41, 0.47, 0.54, 0.60\}$. All curves
			exhibit ultraviolet suppression, confirming that within the
			strong-dispersion regime~$\chi$ modulates the amplitude but not
			the qualitative shape of the spectrum.}
		\label{CurvaturePowerSpectrumVsK2}
	\end{figure}

\begin{figure}[H]
		\centering
		\includegraphics[width=0.45\textwidth]{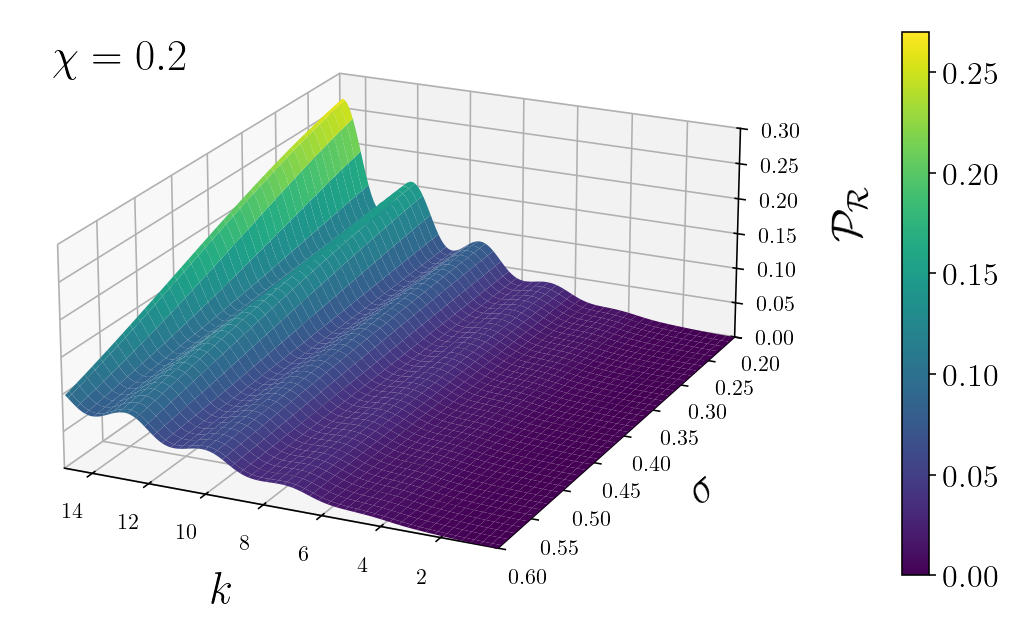}
		\caption{Power spectrum $\mathcal{P}_\mathcal{R}(k,\sigma)$ at
			$\phi = 2$ within Level~2, for fixed $\chi = 0.2$ and $U = 0$.
			The diagonal ridge at $\sigma \approx 0.2$--$0.25$,
			$k \approx 14$--$15$ traces the locus of maximum amplification.
			The spectrum is strongly suppressed toward large~$\sigma$,
			consistent with the transition identified in
			Figure~\ref{CurvaturePowerSpectrumVsK2}.}
		\label{fig_PR_3D_k_sigma}
	\end{figure}

	For $\sigma$ above the critical threshold~$\sigma_c \approx
	0.24$--$0.37$, the friction dominates and the oscillation
	amplitude decreases progressively after the bounce, reflecting
	the physical mechanism in which the gravitational quantum moments
	act as a dissipative channel that drains energy from the scalar
	mode into quantum fluctuations of the background geometry.
	For $\sigma$ below $\sigma_c$, the friction is insufficient to
	overcome the amplification, and the mode amplitude continues
	to grow after the bounce. This conditional behavior is analyzed
	quantitatively in Figures~\ref{CurvaturePowerSpectrumVsK2}%
	--\ref{fig_PR_3D_k_sigma}.

\begin{figure}[H]
		\centering
		\includegraphics[width=0.45\textwidth]{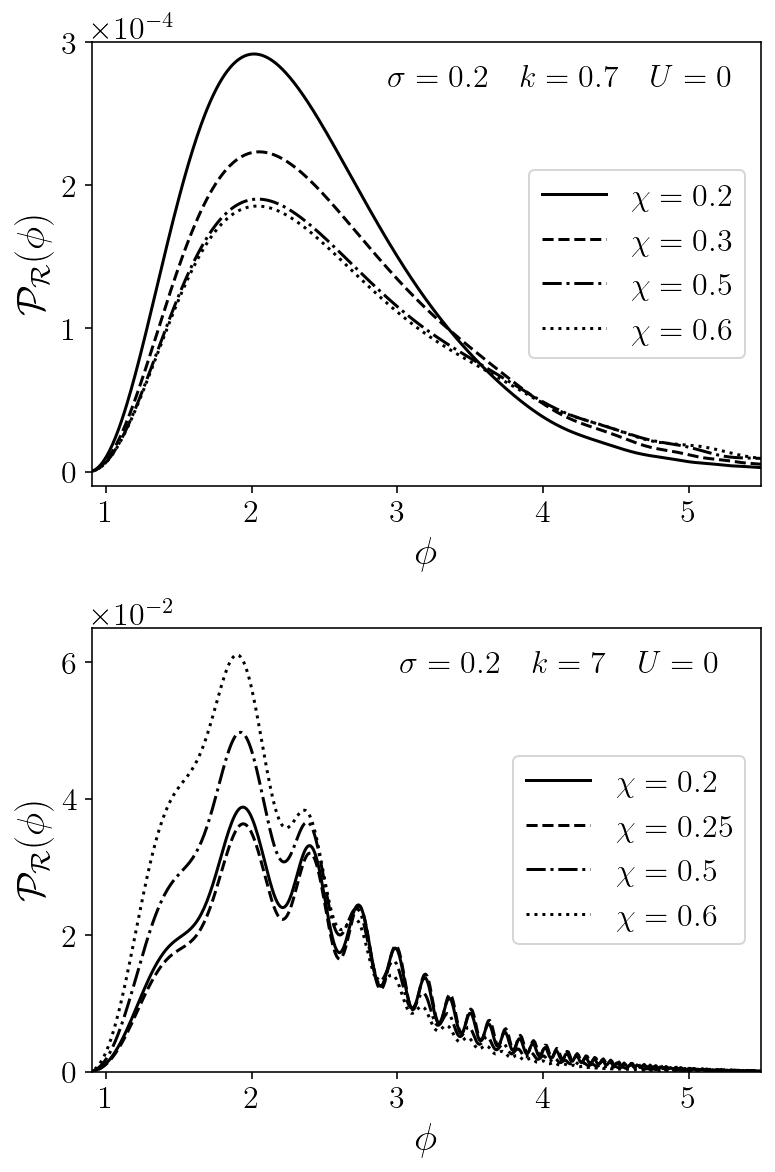}
		\caption{Time evolution of the power spectrum
			$\mathcal{P}_\mathcal{R}(\phi)$ for modes $k=0.7$ (top) and
			$k=8$ (bottom), at fixed $\sigma=0.2$ and various values of the
			scalar dispersion $\chi$. Larger $\chi$ enhances the peak
			amplitude near the bounce ($1\lesssim\phi\lesssim 2$) because
			the initial mode dispersion $G^{vv}_0 = \chi^2$ is larger.
			Post-bounce oscillations are more pronounced for larger $k$ and
			are progressively suppressed as $\chi$ increases, resulting in
			a smoother spectral evolution.}
		\label{CurvaturePowerSpectrumVsTime}
	\end{figure}

	Figure~\ref{CurvaturePowerSpectrumVsTime} shows the time evolution of the power spectrum $\mathcal{P}_\mathcal{R}(\phi)$, computed from Eq.~(\ref{PowerSpectrum}), for two modes, $k = 0.7$ and $k = 7$, and several values of the scalar dispersion $\chi$ at fixed $\sigma$. In both cases, after the bounce (for $\phi \geq 2$), the spectrum decays as the Universe expands, while it exhibits oscillations for larger $k$ values. For low-frequency modes in the bounce region, that is, modes with $k \leq 3$ at $1 \lesssim \phi \lesssim 2$, the power spectrum amplitude is slightly suppressed as $\chi$ increases. This suppression is minimal, since it is of the order of $10^{-4}$; therefore, there is no substantial change in the shape of the spectrum. In contrast, for high-frequency modes in the same region, i.e., $k \geq 4$ at $1 \lesssim \phi \lesssim 2$, the change becomes more evident: the power spectrum is enhanced, and the oscillations are progressively suppressed as $\chi$ increases, resulting in a smoother spectral evolution. This smoothing reflects the dissipative role of the quantum moments: larger quantum fluctuations in the scalar sector reduce the coherence of the post-bounce oscillations.
	
\begin{figure}[H]
		\centering
		\includegraphics[width=0.45\textwidth]{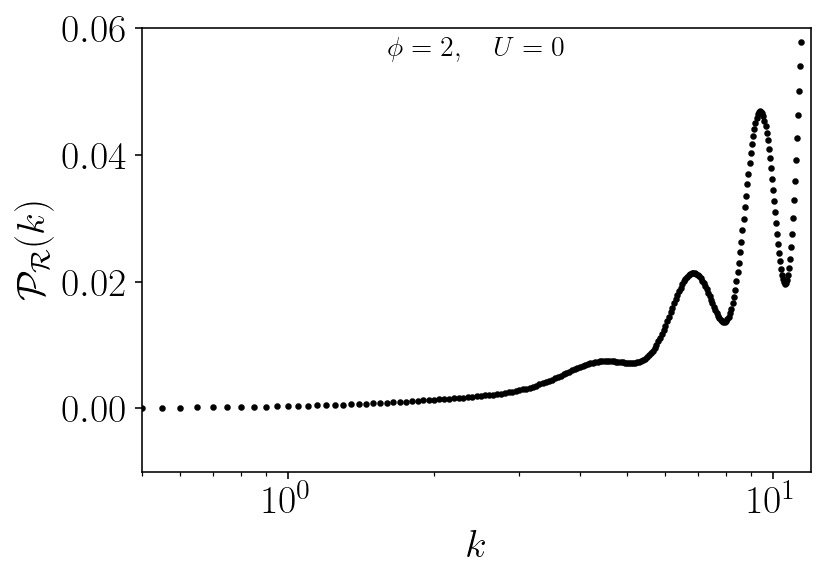}
		\caption{Power spectrum $\mathcal{P}_\mathcal{R}(k)$ at
			$\phi = 2$ without quantum moments.
			The spectrum exhibits large amplitudes, pronounced oscillations,
			and strong ultraviolet growth for $k > 3$.
			Long-wavelength modes ($k < 3$) are only mildly affected
			by the bounce, whereas short-wavelength modes are substantially
			amplified. This ultraviolet behavior is nonphysical and is
			regularized when gravitational quantum moments are included
			(see Figure~\ref{CurvaturePowerSpectrumVsK2}).}
		\label{CurvaturePowerSpectrumVsK1}
	\end{figure}

	It is important to distinguish the role of the two dispersion
	parameters: $\chi$ controls the initial quantum state of the
	scalar mode, while $\sigma$ controls the initial quantum state of the
	gravitational background. Increasing $\chi$ enhances the peak
	power spectrum amplitude because $G^{vv}_0 = \chi^2$ directly
	sets the initial mode dispersion. Increasing $\sigma$, by contrast,
	enhances the gravitational quantum friction $\mathcal{D}$ in
	Eq.~(\ref{DampedFrequencyTerms}) and suppresses the power spectrum after
	the bounce, as shown in Figures~\ref{CurvaturePowerSpectrumVsK1} and~\ref{CurvaturePowerSpectrumVsK2}.
	These two effects are physically distinct and must not be confused.
	%%%%%%%%%%%%%%%
	
	Figures~\ref{CurvaturePowerSpectrumVsK1} and~\ref{CurvaturePowerSpectrumVsK2}
	compare the power spectrum $\mathcal{P}_\mathcal{R}(k)$
	as a function of wavenumber, evaluated at $\phi = 2$
	(immediately after the bounce), within Level~2.
	
	Figure~\ref{CurvaturePowerSpectrumVsK1} shows the pure-bounce
	case (no quantum moments). The spectrum exhibits
	large amplitudes, pronounced oscillatory behavior, and a strong
	growth toward the ultraviolet ($k > 3$). Modes with small
	wavenumber ($k < 3$), corresponding to long wavelengths,
	are only mildly affected by the bounce, while short-wavelength
	modes are substantially amplified. The ultraviolet growth is a
	direct consequence of the bounce dynamics acting most strongly
	on modes whose physical wavelength at the bounce is comparable
	to the bounce scale~$a_{\min}$. In the absence of a
	regularization mechanism, this growth would imply a divergent
	energy density at small scales, which is nonphysical.
	
	Figure~\ref{CurvaturePowerSpectrumVsK2} shows $\mathcal{P}_%
	\mathcal{R}(k)$ at $\phi = 2$ as one-dimensional cuts along~$k$.
	The upper panel fixes $\chi = 0.5$ and shows the dependence on
	the gravitational dispersion $\sigma \in \{0.20, 0.24, 0.31,
	0.37, 0.41, 0.47, 0.54, 0.60\}$. The lower panel fixes
	$\sigma = 0.5$ and shows the dependence on the scalar dispersion
	$\chi \in \{0.20, 0.24, 0.31, 0.37, 0.41, 0.47, 0.54, 0.60\}$.
	
	The upper panel reveals a transition between two qualitatively
	distinct regimes as a function of~$\sigma$:
	\begin{itemize}
		\item \textit{Weak-dispersion regime}
		($\sigma \lesssim 0.24$). The gravitational quantum moments are
		insufficiently large to counteract the bounce-induced
		amplification. The spectrum grows monotonically toward the
		ultraviolet, qualitatively resembling the no-moment case of
		Figure~\ref{CurvaturePowerSpectrumVsK1} but with a moderately
		reduced overall amplitude.
		\item \textit{Strong-dispersion regime}
		($\sigma \gtrsim 0.37$). The gravitational quantum moments
		generate a damping term~$\mathcal{D}$ in
		Eq.~(\ref{EffectiveModeDynamics}) that overcomes the bounce
		amplification. The spectrum develops a peak at
		$k_{\mathrm{peak}} \approx 8$--$10$ and is strongly suppressed
		for $k \gtrsim k_{\mathrm{peak}}$, providing a natural
		ultraviolet regularization of the bounce-induced growth.
	\end{itemize}
	The transition between regimes occurs in the range
	$\sigma_c \approx 0.24$--$0.37$ and reflects the competition
	between the bounce-induced amplification and the dissipative
	friction sourced by the gravitational quantum moments $g_1 \sim
	G^{JJ}$ and $g_2^- \sim G^{J\bar{J}}$ through~$\mathcal{D}$.
	
	The lower panel shows that, the scalar dispersion~$\chi$ modulates the amplitude of the peak, increasing it for modes with $5 < k < 10$; however, for $k > 10$, all curves exhibit ultraviolet suppression. The peak of maximum amplitude shifts from $k_{\mathrm{peak}} \approx 10$ at $\chi = 0.20$ to $k_{\mathrm{peak}} \approx 9$ at $\chi = 0.60$, while the amplitude varies with the initial mode dispersion $G_0^{vv} = \chi^2$. The qualitative shape of the spectrum is set by~$\sigma$, while~$\chi$ acts as an amplitude modulator.
	
\begin{figure}[H]
		\centering
		\includegraphics[width=0.45\textwidth]{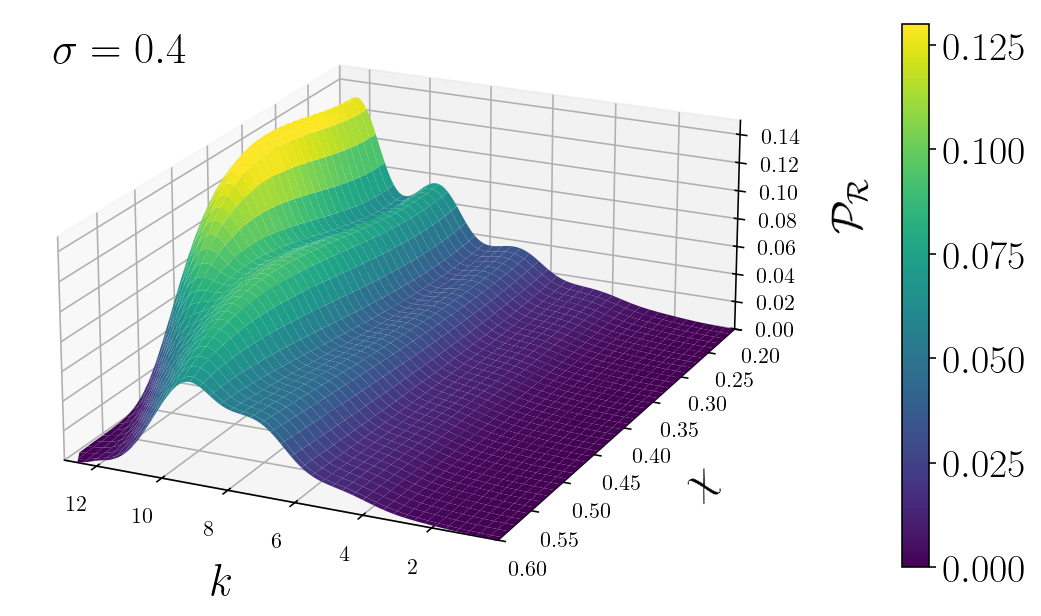}
		\caption{Power spectrum $\mathcal{P}_\mathcal{R}(k,\chi)$ at
			$\phi = 2$ within Level~2, for fixed $\sigma = 0.4$ and $U = 0$.
			The surface displays a clear peak at $k_{\mathrm{peak}} \approx
			9$--$11$ whose amplitude grows as $\chi$ decreases. The
			ultraviolet sector ($k \gtrsim 12$) is strongly suppressed, confirming that the regularization
			mechanism is controlled by the gravitational quantum state.}
		\label{fig_PR_3D_k_chi}
	\end{figure}

\begin{figure}[H]
		\centering
		\includegraphics[width=0.45\textwidth]{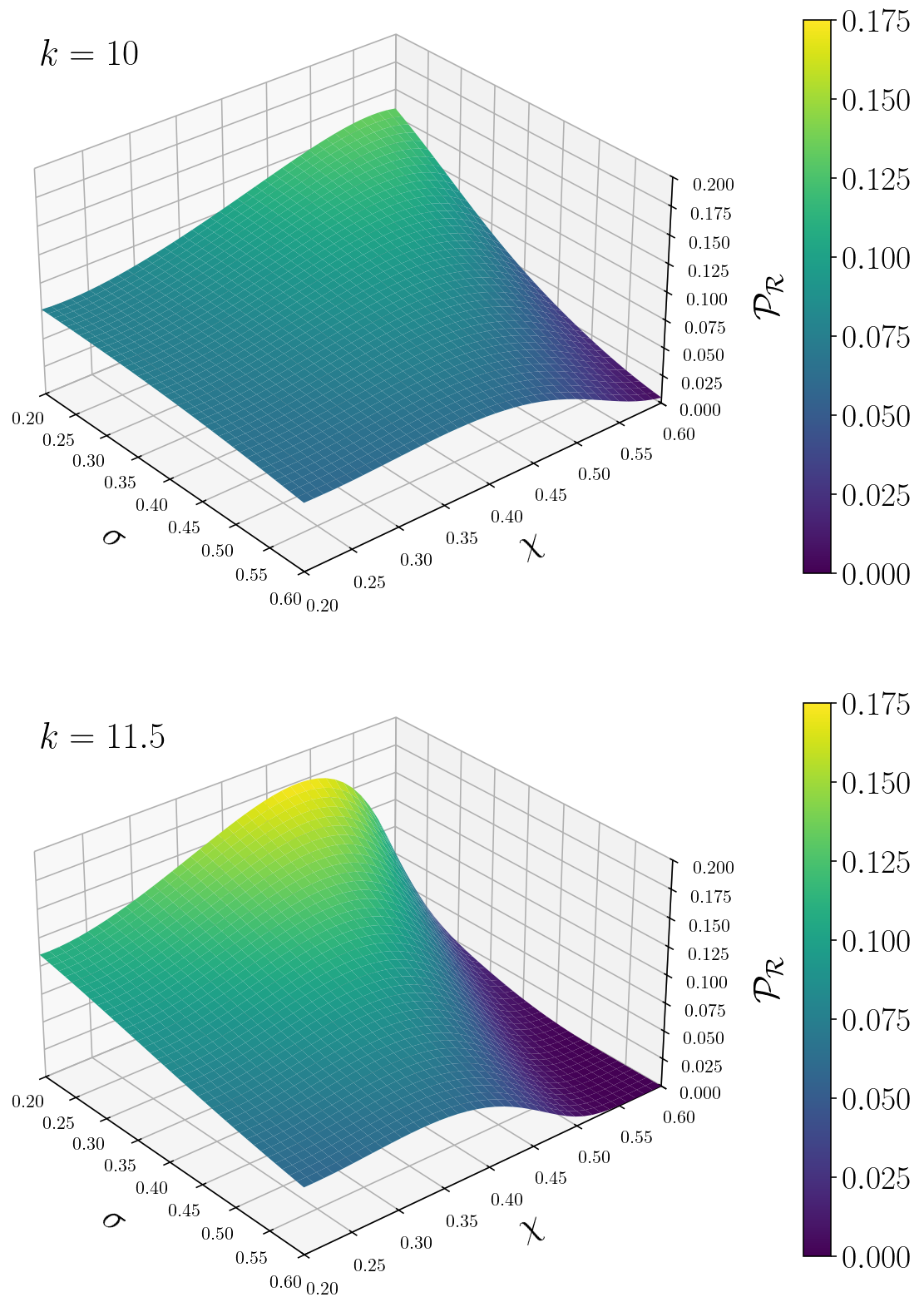}
		\caption{Power spectrum $\mathcal{P}_\mathcal{R}(\sigma,\chi)$
			at $\phi = 2$ within Level~2, for fixed $k = 9$, $k = 11.5$, and $U = 0$.
			The amplitude decreases monotonically with~$\sigma$ (through the
			enhanced gravitational quantum friction~$\mathcal{D}$) and grows
			with $1/\chi$ (through the initial mode dispersion $G_0^{vv} =
			\chi^2$). The separability confirms that~$\sigma$ controls the
			ultraviolet regularization while~$\chi$ acts as an amplitude
			modulator.}
		\label{fig_PR_3D_sigma_chi}
	\end{figure}

	Figures~\ref{fig_PR_3D_k_chi}--\ref{fig_PR_3D_k_sigma} show the
	systematic exploration of the two-dimensional parameter dependence
	of~$\mathcal{P}_\mathcal{R}$ as three-dimensional surfaces.
	
	Figure~\ref{fig_PR_3D_k_chi} shows $\mathcal{P}_\mathcal{R}(k,
	\chi)$ for fixed $\sigma = 0.4$. The surface displays a clear peak
	at $k_{\mathrm{peak}} \approx 9$--$11$ whose amplitude grows as
	$\chi$ decreases, and a clean ultraviolet suppression for
	$k \gtrsim 12$.
	
	Figure~\ref{fig_PR_3D_sigma_chi} shows $\mathcal{P}_\mathcal{R}(\sigma, \chi)$ for fixed $k = 10$ and $k = 11.5$, wavenumbers within the regularized ultraviolet sector. Both surfaces decrease toward the strong quantum regime. The surfaces decrease monotonically with~$\sigma$ and grow with $1/\chi$, confirming that~$\sigma$ controls the suppression and~$\chi$ the overall amplitude.
	
	Figure~\ref{fig_PR_3D_k_sigma} shows $\mathcal{P}_\mathcal{R}(k,
	\sigma)$ for fixed $\chi = 0.2$. The surface displays a diagonal
	ridge tracing the locus of maximum amplification at $\sigma
	\approx 0.2$--$0.25$ and $k \approx 14$--$15$, with clean
	suppression toward large~$\sigma$ for all~$k$.
	
	The existence of the
	threshold~$\sigma_c$ implies that the ultraviolet regularization
	is a conditional rather than universal feature of the second-order
	effective dynamics: it requires that the quantum uncertainty of
	the background geometry be sufficiently large.
	%%%%%%%%%%%%%%%
	\subsection{Comparison with the full system}
	\label{sec:full_system}
	
\begin{figure}[H]
		\centering
		\includegraphics[width=0.38\textwidth]{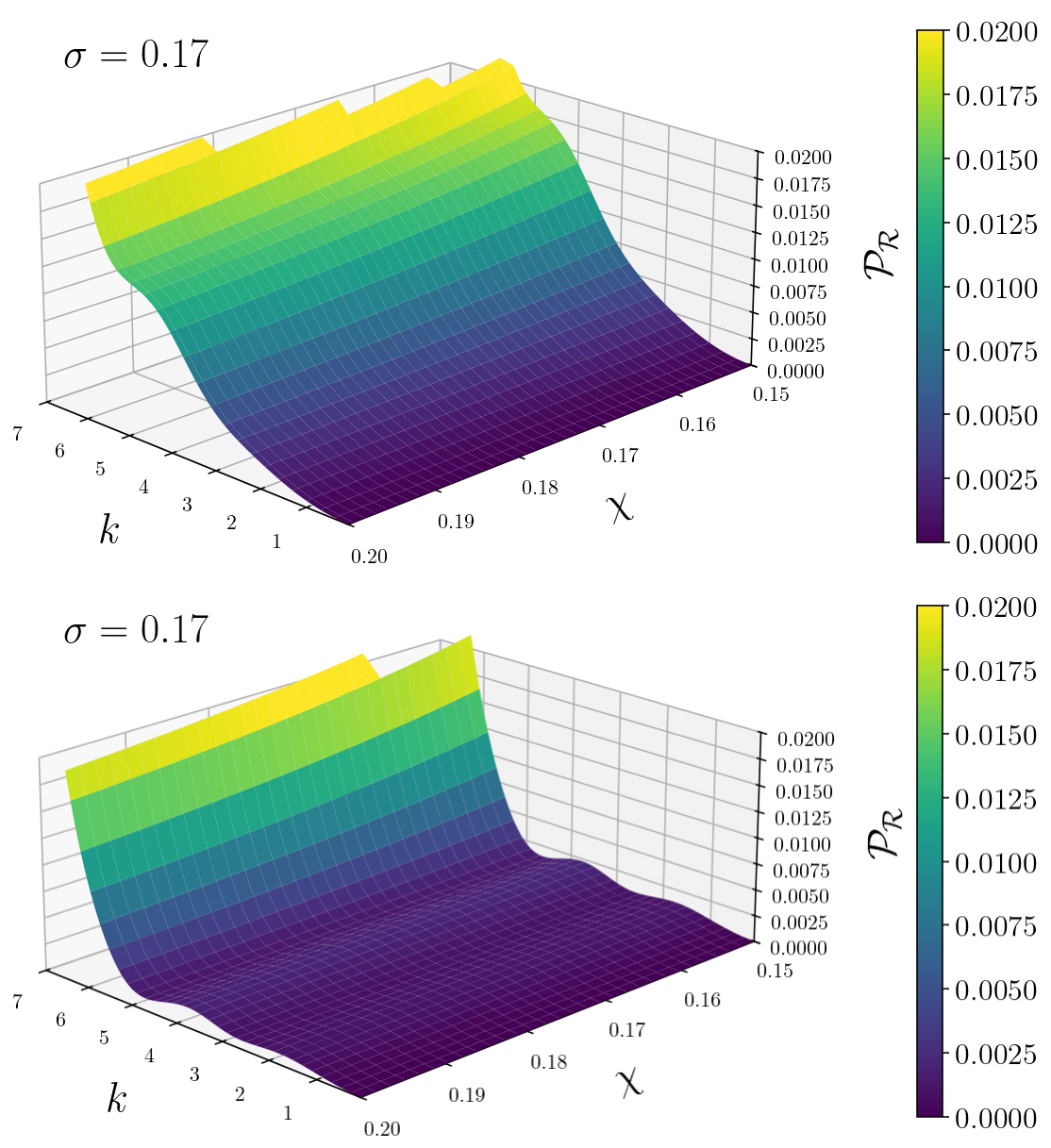}
		\caption{Comparison of the power spectrum
			$\mathcal{P}_\mathcal{R}(k,\chi)$ at $\phi = 2$ between the
			full second-order system (Level~3, lower panel) and Level~2
			(upper panel), at fixed $\sigma = 0.17$ and $U = 0$.
			Both levels share the same qualitative spectral
			structure---peak followed by ultraviolet growth---but Level~3
			produces systematically smaller amplitudes ($\mathcal{P}_
			\mathcal{R} \approx 0.05$ vs.\ $0.175$). The cross-sector
			quantum moments $G^{Jv}, G^{J\pi}, G^{pv}, G^{p\pi}$ strongly suppress the spectrum for $k<5$, while enhanced it exponentially for $k>5$.}
		\label{fig_PR_3D_comparison}
	\end{figure}

\begin{figure}[H]
		\centering
		\includegraphics[width=0.35\textwidth]{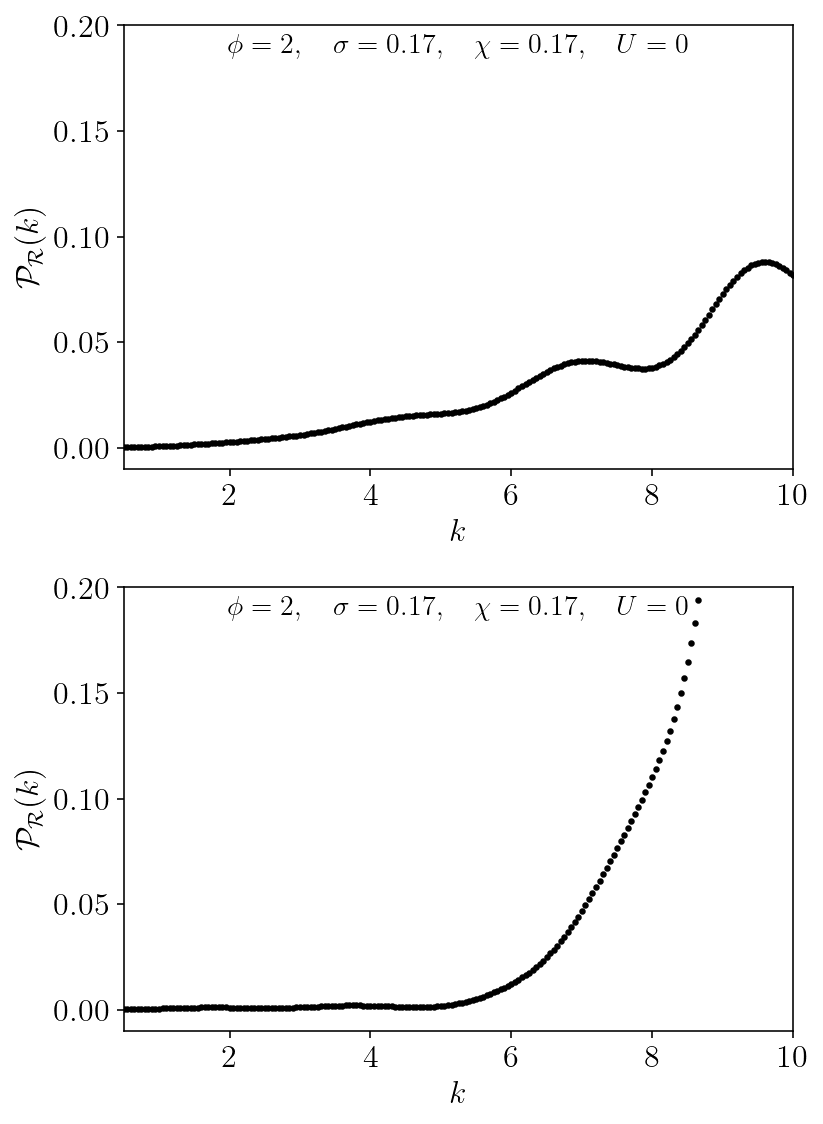}
		\caption{Comparison of $\mathcal{P}_\mathcal{R}(k)$ at
			$\phi = 2$ between Level~3 (lower panel) and Level~2 (upper
			panel), at fixed $\sigma = 0.17$, $\chi = 0.17$, $U = 0$. The figure shows the exponential growth of $\mathcal{P}_\mathcal{R}(k)$ once the cross-correlation terms have been included.}
		\label{fig_PR_2D_comparison}
	\end{figure}

	We now compare the Level~2 results discussed above with those
	obtained from the full second-order system (Level~3), in which
	the cross-sector quantum moments $G^{Jv}, G^{J\pi}, G^{pv},
	G^{p\pi}$ are also evolved through equations
	$G_7$--$G_{10}$ rather than being set to
	zero. Figures~\ref{fig_PR_3D_comparison}
	and~\ref{fig_PR_2D_comparison} show this comparison at fixed
	$\sigma = 0.17$, in the weak-dispersion regime.
	
	Figure~\ref{fig_PR_3D_comparison} shows the surface
	$\mathcal{P}_\mathcal{R}(k, \chi)$ at $\sigma = 0.17$ for the
	full system (lower panel) and for Level~2 (upper panel). The
	qualitative structure of the spectrum---a peak at intermediate
	wavenumbers with growth toward the ultraviolet---is shared by
	both levels, indicating that the cross-sector moments do not
	introduce qualitatively new spectral features. However, the
	amplitude of the spectrum is systematically smaller at Level~3 for modes of lower frequency $k<5$ ,
	with the surface reaching values up to $\mathcal{P}_\mathcal{R}
	\approx 0.05$ at $k\approx6$ compared to $\mathcal{P}_\mathcal{R} \approx 0.175$
	at Level~2. However, the cross-sector moments for $k>5$ amplify exponentially the spectrum relative to the truncation in which they are absent.
	
	Figure~\ref{fig_PR_2D_comparison} shows one-dimensional cuts of Fig.~\ref{fig_PR_3D_comparison} at $\sigma = 0.17$ and $\chi = 0.17$ for reference. In both figures ~\ref{fig_PR_3D_comparison} and \ref{fig_PR_2D_comparison}, the exponential growth that appears once the cross-correlation terms are included may be associated with gravitational instabilities arising from the truncation order of the effective system.
	
	The numerical instabilities of the full second-order system in
	the ultraviolet sector are physically informative. They indicate
	that the second-order truncation of the moments hierarchy, while
	consistent at the formal level of the $\hbar$-expansion, becomes
	insufficient for high-wavenumber modes whose dynamics involves
	significant cross-sector correlations. A complete treatment of the
	ultraviolet sector would require either the inclusion of
	higher-order moments or alternative resummation strategies. The
	Level~2 approximation can be understood as a controlled truncation
	that retains the dominant effect of the gravitational quantum
	moments on the perturbation mode while suppressing the
	backreaction channel responsible for the instabilities. The
	qualitative agreement between the two levels for $k \lesssim 7$
	supports the use of Level~2 as the natural regime from which the
	conditional ultraviolet regularization of
	Figures~\ref{CurvaturePowerSpectrumVsK2}--\ref{fig_PR_3D_k_sigma}
	can be reliably extracted.
	%%%%%%%%%%%%%%%%%%%%%%%%%%
	
	\section{Conclusions and discussion}\label{sec:conclusions}	
	In this work we have computed second-order quantum corrections, in the sense of quantum dispersions and correlations (quantum moments), to a cosmological model coupling a single scalar perturbation mode to a bouncing background within Loop Quantum Cosmology. Two complementary analyses have been carried out. In the first, we adopt the test-field approximation, in which the scalar mode propagates on a fixed semiclassical background without backreaction, and derive an analytical correction to the primordial power spectrum in the de~Sitter regime. In the second, we solve the full second-order effective dynamics numerically in the vicinity of the cosmic bounce with vanishing scalar potential $U=0$, retaining the coupling between the gravitational and scalar quantum moments.
	
	\subsection{Description of the work}
	We start from a Hamiltonian formulation that couples the
	gravitational background to the scalar sector through an effective
	frequency expressed in terms of phase-space variables. The
	deparametrization of this system using the scalar field $\phi$
	as internal time results in the quartic polynomial
	equation~(\ref{FourPolynomicalEquation}), whose solution yields the
	deparametrized Hamiltonians~(\ref{ClassicalHamiltonian}) under the
	assumptions of a constant scalar field potential $U$ and a small
	perturbation amplitude $v_k/a \ll 1$. These conditions restrict the
	analysis to scalar modes propagating in a regime close to the
	initial singularity. The cosmic bounce is then incorporated by
	introducing holonomy corrections within the $\mu_0$ scheme of LQC,
	implemented through the non-canonical
	variables~(\ref{eq:JJbar}). The effective quantization
	method of Refs.~\cite{bojowald2006effective,bojowald2012quantum} is applied
	to obtain the second-order effective Hamiltonian~(\ref{SecondOrderHamiltonian}),
	which extends the classical phase space by treating quantum
	dispersions and correlations as additional dynamical degrees of
	freedom. The resulting second-order effective dynamics is given by
	equations~(\ref{SecondOrderEOM1}) and~(\ref{SecondOrderEOM2}). In the absence
	of the scalar perturbative sector, this dynamics reduces
	consistently to that of Ref.~\cite{bojowald2007effective}.
	
	Within the test-field approximation, the effective scale factor
	exhibits a smooth bounce described by~(\ref{backgroundSolutions}),
	and the dynamics of the scalar mode satisfies
	equation~(\ref{etaSimplfiedMode}), which takes the form of a
	harmonic oscillator with a time-dependent frequency and an
	effective damping term. The damping coefficient $\mathcal{D}$
	in~(\ref{EffectiveModeDynamics}) is sourced by the gravitational quantum
	moments $g_1$ and $g_2^-$ and encodes the dissipative effect of
	background quantum fluctuations on the perturbation mode. Through
	the variable change $v_k = \sqrt{a\mathcal{F}}\,u_k$, the mode equation
	acquires the structure of a modified Mukhanov--Sasaki
	equation~(\ref{ModifiedMukhanovEquation}) with LQC-corrected background
	functions. This equation reduces to the standard Mukhanov--Sasaki
	form~(\ref{eq:MS}) when $\mathcal{F} = -1/a$, which is recovered in the
	classical holonomy limit $\mu c\to 0$ in the de~Sitter
	regime (Section~\ref{subsec:classicallimit}).
	
	The equations of motion for the quantum moments of the scalar sector
	combine into the single third-order
	equation~(\ref{ThirdOrderEq2}) for the mean squared dispersion
	$G^{vv}$ of the Mukhanov--Sasaki variable. In the limit $\mathcal{F}=-1/a$,
	this equation reduces to that of Ref.~\cite{brizuela2019moment}, providing
	a consistency check. Its solution in the de~Sitter bouncing
	background, obtained perturbatively by treating the LQC correction
	as a small departure from the singular case, yields the corrected
	dimensionless curvature power
	spectrum~(\ref{PowerSpectrumCorrection}). The leading correction is
	proportional to $\ell_{\mathrm{Pl}}^6$ and produces a
	scale-dependent enhancement
	$\delta\mathcal{P}_\mathcal{R} \propto (k\ell_{\mathrm{Pl}})^6$,
	with an associated spectral tilt modification
	$\delta n_s \sim 6C(k\ell_{\mathrm{Pl}})^6 \ll 1$ for all
	cosmologically observable modes (Section~\ref{sec:dS_bounce}).
	This is fully consistent with the Planck measurement
	$n_s = 0.965 \pm 0.004$. The absence of the discretisation
	parameter $\mu$ in~(\ref{PowerSpectrumCorrection}) is a consequence of the
	de~Sitter approximation: $\mu$ drops out because the holonomy
	correction reduces to the classical connection at late times.
	In more general scenarios such as the slow-roll approximation,
	$\mu$ would appear explicitly; its absence here does not indicate
	a lack of quantum corrections but rather that the bounce
	information is encoded in the initial conditions of the de~Sitter
	phase.
	%%%%%%%%%%%%%%%%%%%%%%%%%%%%%
	
	The second-order system is solved numerically near the bounce with
	$U = 0$ (Section~\ref{sec:numerical}), within the Level~2
	approximation in which the gravitational quantum moments are
	dynamically evolved while the cross-sector moments $G^{Jv},
	G^{J\pi}, G^{pv}, G^{p\pi}$ are kept at zero. The background
	evolution exhibits a smooth bounce in the region
	$0 \lesssim \phi \lesssim 2$
	(Figure~\ref{Bounce}), consistent with the analytical
	solution~(\ref{ScaleFactorEvolution}). The scalar modes oscillate
	during contraction, freeze near the bounce, and resume oscillation
	during expansion. The amplitude evolution depends critically on
	the gravitational dispersion~$\sigma$: above a threshold
	$\sigma_c \approx 0.24$--$0.37$, the damping term~$\mathcal{D}$
	in~(\ref{EffectiveModeDynamics}), sourced by the gravitational
	quantum moments, drains energy from the scalar mode during the
	expansion phase and suppresses the ultraviolet sector of the power
	spectrum. Below this threshold, the bounce-induced amplification
	dominates and the ultraviolet growth persists. The peak amplitude
	scales with the scalar dispersion~$\chi$ as expected from
	$G_0^{vv} = \chi^2$, while the peak location and the suppression
	scale are controlled by~$\sigma$
	(Figures~\ref{CurvaturePowerSpectrumVsK2}--\ref{fig_PR_3D_k_sigma}).
	
	A comparison with the full second-order system (Level~3, in which
	the cross-sector moments are also evolved) reveals that the
	qualitative structure of the spectrum---peak followed by
	ultraviolet suppression---is preserved
	(Figures~\ref{fig_PR_3D_comparison} and~\ref{fig_PR_2D_comparison}).
	However, the cross-sector moments amplify the spectrum
	amplitude relative to Level~2 and trigger numerical instabilities
	at high wavenumbers, indicating that the second-order truncation
	of the moments hierarchy is insufficient in the ultraviolet regime.
	The Level~2 approximation can be understood as a controlled
	truncation that captures the dominant effect of gravitational
	quantum fluctuations on the perturbation mode while suppressing
	the backreaction channel responsible for the instabilities.
	
	The analytical enhancement of~(\ref{PowerSpectrumCorrection}) and
	the conditional numerical suppression seen in
	Figures~\ref{CurvaturePowerSpectrumVsK2}--\ref{fig_PR_3D_k_sigma}
	are complementary, not contradictory.
	Equation~(\ref{PowerSpectrumCorrection}) is derived within
	Level~1 at zeroth order in gravitational quantum moments and
	captures the kinematic imprint of the bounce on~$G^{vv}$ at the end of a de Sitter regime.
	The conditional suppression arises from the dynamical evolution of
	the gravitational moments at Level~2 near the bounce regime and represents a new physical
	effect that is genuinely absent in the analytical estimate. Both
	effects are present in the complete theory and operate in different
	sectors of the approximation hierarchy.
	
	\subsection{Physical feasibility and motivation}
	Despite the successes of the standard theory of cosmological
	perturbations and the inflationary scenario in explaining the
	near-scale-invariant power spectrum and the acoustic peaks in the
	CMB~\cite{hu1996small,Planck2018params}, several anomalies at large angular scales remain
	unexplained, including the suppression of power at low multipoles,
	dipolar asymmetry, and a preference for odd-parity
	correlations~\cite{agullo2021large, schwarz2016cmb, sanchis2022missing}.
	These observations have generated significant interest in
	cosmological perturbation models on nonsingular backgrounds,
	which may provide physical mechanisms beyond simple slow-roll
	inflation.  A systematic program within the hybrid quantization
	approach has explored whether the pre-inflationary dynamics of
	LQC can address these anomalies, with detailed analyses of the
	power spectrum~\cite{agullo2013pre} and comprehensive studies of the
	interplay between Planck-scale physics and large-angle CMB
	features~\cite{ashtekar2021cosmic}.  Furthermore, a cosmic bounce opens the
	possibility of connecting LQC with observational data, allowing
	the theory to be directly tested against CMB measurements.
	
	However, bounce models that attempt to resolve the low-multipole
	anomalies through large non-Gaussianities face stringent
	constraints.  In~\cite{van2023constraining}
	it was demonstrated that the values of $|f_{\mathrm{NL}}|$ required
	to mitigate the large-scale power suppression are excluded at high
	significance by the Planck bispectrum data, even though the
	corresponding bispectrum decays exponentially on sub-horizon scales.
	This result highlights an important complementarity with the
	present work: while the bispectrum constraints rule out large
	bounce-induced non-Gaussianities at the three-point level, our
	calculation shows that the leading correction to the two-point power
	spectrum is suppressed by $(k\ell_{\mathrm{Pl}})^{6}$ and is
	therefore negligible at all observable scales.
	The corrections computed here do not provide a mechanism for
	the observed low-multipole power suppression, which would require
	either nontrivial pre-bounce initial conditions, additional quantum
	backreaction effects at higher order in the moments expansion,
	the inclusion of inflationary dynamics beyond the de~Sitter
	approximation, or a careful assessment of the role of the
	regularization scheme~\cite{kowalczyk2025primodial}.  Addressing these anomalies
	remains an important open problem that motivates the future
	extensions listed in Section~VI\,C.
	
	Although the model developed in this work is restricted to the
	$\mu_0$ scheme and to a controlled truncation of the moments
	hierarchy, it captures essential features of bounce dynamics
	and their impact on scalar cosmological perturbations, including
	the analytical correction to the power spectrum and the
	identification of a conditional dynamical regularization
	mechanism: when the gravitational dispersion exceeds a critical
	threshold $\sigma_c \approx 0.31$--$0.37$, the quantum
	fluctuations of the background geometry suppress the ultraviolet
	sector of the perturbation spectrum, while below this threshold
	the bounce-induced amplification dominates.
	In this sense, it provides
	a well-defined framework for exploring potential observational
	signatures of quantum gravitational effects.
	
	Regarding the scheme dependence of the model: our analysis employs
	the reduced Ashtekar variables $(c,p)$ in the $\mu_0$ scheme, in
	which the discretization parameter is held constant. The improved
	$\bar\mu$ scheme of Ref.~\cite{ashtekar2006quantum} is known to provide
	more physically consistent semiclassical behavior by eliminating
	spurious quantum effects at large scales. In principle, the
	dynamics obtained in the $\bar\mu$ scheme would differ
	from~(\ref{SecondOrderEOM1}) and~(\ref{SecondOrderEOM2}). However, for the specific de~Sitter background considered in
	Section~IV\,C, the improved scheme modifies only the numerical
	coefficients of the quantum correction in~(75), without altering
	its functional form or the qualitative physical effects reported
	here.  This expectation is supported by the recent analysis
	of Ref.~\cite{kowalczyk2025primodial}, which provides a systematic comparison
	of the primordial power spectrum across different LQC
	regularization schemes and confirms that, while the detailed
	spectral features, in particular the amplitude of
	bounce-induced oscillations and the location of spectral
	peaks, are scheme-dependent, the qualitative structure of
	the corrections and their suppression at sub-Planckian scales
	remain robust.  A full reformulation of the present calculation
	in the~$\bar\mu$ scheme is listed as a direction for future work.
	
	Finally, the present analysis is formulated within an effective framework, since a full symmetry-unreduced quantum theory is currently unknown. However, a natural strategy to explore its possible phenomenological consequences is to start from the classical theory and consistently incorporate quantum corrections while requiring that the classical theory be recovered in the appropriate limit. In this direction, our work should be regarded as an effective description that reproduces the standard equations of General Relativity in the combined semiclassical and low-curvature limits, namely, $\hbar\rightarrow0$ and $\mu\rightarrow0$. Similar effective strategies have been successfully employed in other approaches to cosmological perturbations in loop quantum cosmology, such as the hybrid quantization and dressed metric frameworks \cite{elizaga2021hybrid, agullo2012quantum, bolliet2015comparison}.
	
	\subsection{Future work}
	\label{sec:future}
	The results presented here are based on a set of physically
	motivated approximations. Several important aspects remain to be
	explored:
	\begin{enumerate}
		\item \textit{Backreaction and higher-order moments.}
		The numerical instabilities of the full second-order system at
		high wavenumbers
		(Figures~\ref{fig_PR_3D_comparison} and~\ref{fig_PR_2D_comparison})
		indicate that the second-order truncation of the moments hierarchy
		is insufficient in the ultraviolet sector. Including third-order
		and higher moments would provide a controlled stabilization of the
		truncation and a systematic study of gravitational backreaction on
		the scalar perturbation spectrum, potentially modifying the power
		spectrum correction in~(\ref{PowerSpectrumCorrection}) at higher
		order. In particular, the threshold $\sigma_c$ identified in
		Section~\ref{sec:numerical} for the onset of ultraviolet
		regularization should be investigated analytically to determine
		whether it has a universal character or depends on the specific
		form of the Hamiltonian and the truncation order.

		\item \textit{Tensor sector.} Including tensor perturbations
		within the same effective moments framework would yield a
		prediction for the tensor power spectrum and the tensor-to-scalar
		ratio $r$. Comparing with the Planck and BICEP/Keck constraints
		on $r$ would provide a direct observational test of the model. We are currently
		analyzing this scenario.
		
		\item \textit{Beyond de~Sitter: slow-roll inflation and dynamical
			potential.} The de~Sitter background and vanishing potential
		adopted here are the simplest setting in which the calculation
		is tractable. Extending the analysis to a slow-roll inflationary
		background with a dynamical potential $U(\phi)$ would make
		the results more directly comparable with CMB observations and
		would restore the explicit $\mu$-dependence in the power spectrum.
		
		\item \textit{Improved $\bar\mu$ scheme.}
		Reformulating the model using the improved
		Ashtekar--Barbero variables of the~$\bar\mu$
		scheme~\cite{ashtekar2006quantum} would eliminate the large-scale
		quantum artifacts of the~$\mu_0$ scheme and place the
		semiclassical analysis on firmer physical grounds.
		The systematic comparison of regularization schemes
		carried out in Ref.~\cite{kowalczyk2025primodial} provides a natural
		benchmark for such an extension and suggests that the
		functional form of the power spectrum correction~(75)
		is likely preserved, with modifications confined to
		numerical coefficients and the detailed structure of the
		spectral oscillations near the bounce scale.
		
		\item \textit{Non-Gaussian initial states.} The Gaussian initial
		conditions employed here are the simplest physically motivated
		choice. Exploring non-Gaussian initial quantum states could
		reveal additional effects of the quantum geometry on the
		primordial power spectrum and non-Gaussianities.
		
		\item \textit{Third-order moments and the bispectrum.}
		Extending the effective moments expansion to third order
		would yield predictions for the primordial bispectrum within
		the present framework, enabling a direct comparison with the
		observational constraints of Ref.~\cite{van2023constraining} and with
		the non-Gaussianity calculations performed within the hybrid
		quantization approach~\cite{agullo2021large}.  Given that van~Tent
		\textit{et~al.} have shown that bouncing models with large
		$|f_{\mathrm{NL}}|$ are excluded by Planck, establishing
		whether the moments formalism produces comparably large or
		naturally suppressed non-Gaussianities would constitute an
		important consistency test of the framework.
		
		\item \textit{Multi-mode extension.} The present analysis has been restricted to a single scalar perturbation mode in order to analyze the effects of the LQC cosmic bounce on the perturbative sector while keeping the resulting effective dynamics analytically and numerically tractable. A natural extension of the present formalism is to incorporate several perturbation modes, following the approach of~\cite{brizuela2021mode}. Extending the present framework in this direction would provide a more realistic description of the effects of the LQC cosmic bounce on cosmological perturbations and would allow the study of correlations among different perturbation modes.
	\end{enumerate}
	
	\section{ACKNOWLEDGMENTS}
	G.S.H. acknowledges the financial support provided by SECIHTI through a doctoral scholarship. H.H.H. acknowledges SECIHTI Sabbatical Grant 2025. H.A.M.T. acknowledges the support of SNII-14585. This work was supported by CONAHCYT/SECIHTI Grant CBF-2023-2024-1937. We thank J. Arroyo and C. Javier for helpful discussions. 
	\section{DATA AVAILABILITY}
	
	The data that support the findings of this article are openly available \cite{SanchezHerrera2026Data}.
	\begin{widetext}
		\appendix
		\section{Homogeneous sector in the effective constraints approach}
		\label{AppendixEffectiveConstraints}
		
		In this appendix, we illustrate the connection between our approach and the effective constraints formalism, described in detail in \cite{bojowald2009effective}, by considering the homogeneous sector of the Hamiltonian constraint while neglecting perturbations for simplicity.
		
		We start with the homogeneous sector of Eq.~(\ref{GravitationalMatterSingleScalarHamiltonian}), which reads
		\begin{eqnarray}\label{HomogeneousConstraint}
			H_{\text{M}}^{\text{h}}  &=&  N \left[-\frac{l^{2}P_{a}^{2}}{4 a \mathcal{V}} + \frac{P_{\phi}^{2}}{2 a^{3}\mathcal{V}} + \frac{a^{3}\mathcal{V} U}{2} \right] \approx 0.
		\end{eqnarray}
		For simplicity, we set the fiducial spatial volume $\mathcal{V}=1$, and rewrite the constant $l^{2}=2 G$, where $G:= 4\pi G_{N}/3$, as well as $\tilde{P}_{\phi}=P_{\phi}/\sqrt{2}$ and $\tilde{U}=U/2$. Applying these simplifications to (\ref{HomogeneousConstraint}), we define the Hamiltonian constraint $C_{\text{H}}$ as follows
		\begin{eqnarray}
			C_{\text{H}} &=& - \frac{G}{2}\frac{P_{a}^{2}}{a} + \frac{\tilde{P}_{\phi}^{2}}{a^{3}} +a^{3}\tilde{U} \approx 0,
		\end{eqnarray}
		which in terms of the Ashtekar-Barbero variables $|p|=a^{2}$ and $c=\gamma\dot{a}$ is given by
		\begin{eqnarray}\label{Hconstraint|p|}
			C_{\text{H}} &=& -\frac{c^{2}\sqrt{|p|}}{2\tilde{G}}+|p|^{-3/2}\tilde{P}_{\phi}^{2}+|p|^{3/2} \tilde{U} \approx 0,
		\end{eqnarray}
		where $\tilde{G}=\gamma^{2} G$. Including the bounce effects in the above equation through the holonomy correction (\ref{eq:JJbar}), and rescaling the Hamiltonian constraint $C_{\text{H}}$, we obtain the following equation
		\begin{eqnarray}
			C_{\text{H}}' &:=& p^{3/2} C_{\text{H}} =\frac{(J-\bar{J})^{2}}{8 \tilde{G} }+\tilde{P}_{\phi}^{2}+p^{3}\tilde{U} \approx0.
		\end{eqnarray}
		Following the effective constraint approach, we need to calculate the infinite set of effective constraints through the equation
		\begin{eqnarray}
			\left\langle  (\hat{J}-\langle \hat{J} \rangle)^{i}(\hat{J}^{\dagger}-\langle \hat{J}^{\dagger} \rangle)^{j}(\hat{p}-\langle \hat{p}\rangle)^{k}(\hat{\phi}-\langle\hat{\phi}\rangle)^{m}(\hat{\tilde{P}}_{\phi}-\langle \hat{\tilde{P}}_{\phi}\rangle)^{n}\hat{C}'_{\text{H}}\right\rangle =0,
		\end{eqnarray}
		which at second order are summarized by the following six effective constraints
		\begin{eqnarray}
			C_{\mathcal{H}}&=&  C_{\text{H}}'+\frac{1}{8 \tilde{G}}\left(G^{JJ} 
			- 2G^{J\bar{J}} 
			+ G^{\bar{J}\bar{J}} \right) 
			+ G^{\tilde{P}_{\phi}\tilde{P}_{\phi}}   		 
			+ 3 G^{pp}  p  \tilde{U} 
			+ 3 G^{p\phi}  p^2 \tilde{U}' 
			+ \frac{1}{2} G^{\phi\phi}  p^3 \tilde{U}''    \approx 0,  \label{Constraint1} \\
			C_{J} &=& \frac{(J - \bar{J})}{8 \tilde{G} } \left( G^{J J} - G^{J \bar{J}} - \hbar p - \frac{\hbar^2}{2} \right) +\tilde{P}_{\phi}G^{J \tilde{P}_{\phi}} 
			+ \frac{3 p^2 \tilde{U}}{2 } \left( G^{J p} + \frac{\hbar}{2} J \right)
			+ \frac{p^3 \tilde{U}'}{2 } G^{J \phi} \approx 0 , \label{Constraint2}\\ 
			C_{\bar{J}} &=& \frac{(J - \bar{J})}{8 \tilde{G} } \left( G^{J \bar{J}} + \hbar p + \frac{\hbar^2}{2} - G^{\bar{J} \bar{J}} \right) +\tilde{P}_{\phi}G^{\bar{J}\tilde{P}_{\phi}}
			+ \frac{3 p^2 \tilde{U}}{2 } \left( G^{\bar{J} p} + \frac{\hbar}{2} J \right)
			+ \frac{p^3 \tilde{U}'}{2} G^{\bar{J} \phi}\approx0 , \label{Constraint3}\\ 
			C_{p} &=&  \frac{(J - \bar{J})}{8 \tilde{G} } \left( G^{J p} - G^{\bar{J} p} + \frac{\hbar}{2} (J + \bar{J}) \right) +\tilde{P}_{\phi}G^{p\tilde{P}_{\phi}}
			+ \frac{3 p^2 \tilde{U}}{2 } G^{p p}
			+ \frac{p^3 \tilde{U}'}{2 } G^{p \phi} \approx0 , \label{Constraint4}\\ 
			C_{\phi} &=&  \frac{(J - \bar{J})}{8 \tilde{G} } \left( G^{J \phi} - G^{\bar{J} \phi} \right)  +\tilde{P}_{\phi}\left(G^{\phi \tilde{P}_{\phi}} -\frac{i \hbar}{2}\right)
			+ \frac{3 p^2 \tilde{U}}{2 } G^{p \phi}
			+ \frac{p^3 \tilde{U}'}{2 }G^{\phi \phi} 
			\approx0 , \label{Constraint5}\\ 
			C_{\tilde{P}_{\phi}} &=&   \frac{(J - \bar{J})}{8 \tilde{G}} \left( G^{J \tilde{P}_{\phi}} - G^{\bar{J} \tilde{P}_{\phi}} \right) +\tilde{P}_{\phi}G^{\tilde{P}_{\phi} \tilde{P}_{\phi}}
			+ \frac{3 p^2 \tilde{U}}{2 } G^{p \tilde{P}_{\phi}}
			+ \frac{p^3 \tilde{U}'}{2 } \left( G^{\phi \tilde{P}_{\phi}} + \frac{i \hbar}{2} \right) \approx0  . \label{Constraint6} 
		\end{eqnarray}
		Here the primes in the potential $\tilde{U}$ denotes derivatives respect to the scalar field $\phi$. We now deparameterize the system by choosing the scalar field $\phi$ as the internal time, so that its conjugate momentum $\tilde{P}_{\phi}$ plays the role of the physical Hamiltonian. To this end, we first use the effective constraints
		(\ref{Constraint2})--(\ref{Constraint5}) to solve for the quantum moments involving $\tilde{P}_{\phi}$. These expressions are then substituted into Eq.~(\ref{Constraint6}), yielding $G^{\tilde{P}_{\phi}\tilde{P}_{\phi}}$ in terms of the remaining expectation values and quantum moments. Finally, inserting this result into the Hamiltonian constraint (\ref{Constraint1}) leads to a closed equation for $\tilde{P}_{\phi}$,
		\begin{eqnarray}\label{Generalfourthorderpiequation}
			64 \tilde{G}^{2}\tilde{P}_{\phi}^{4}+ a_{1} 8 \tilde{G}	 \tilde{P}_{\phi}^{2}+ a_{2} \tilde{P}_{\phi}+a_{3} \approx 0,
		\end{eqnarray}
		where the coeficients $a_{0}$, $a_{1}$ and $a_{2}$, are to first order in $\hbar$,

		\begin{eqnarray}\label{coefficientsa123}
			a_{1} &=&  (J - \bar{J})^2+ \left(G^{J J} - 2 G^{J \bar{J}} + G^{\bar{J} \bar{J}}\right) + 4 \tilde{G} p \left[ 2 (3 G^{p p} + p^2) \tilde{U} + p \left( 6G^{p \phi} \tilde{U}' + G^{\phi \phi} p \tilde{U}'' \right)\right],
			\nonumber \\
			a_{2}&=&   16 \, i \hbar \, \tilde{G}^2   p^3 \tilde{U}', \nonumber \\ 
			a_{3} &=&  (J - \bar{J})^2 \left( G^{J J} - 2 G^{J \bar{J}} + G^{\bar{J} \bar{J}} -  2\hbar p \right) \nonumber \\
			&& + 2 \tilde{G} p^2 \left[
			3 \tilde{U} (J - \bar{J}) \left( 4 G^{J p} - 4 G^{\bar{J} p} + 3 \hbar J + \hbar \bar{J} \right)
			+ 24 \tilde{G} G^{p p} p^2 \tilde{U}  \right. \nonumber \\
			&&\left.
			+ 4 p \left( 
			(G^{J \phi} - G^{\bar{J} \phi})(J - \bar{J}) 
			+ 12 \tilde{G} G^{p \phi} p^2 \tilde{U} 
			\right) \tilde{U}' 
			+ 8 \tilde{G} G^{\phi \phi} p^4 (\tilde{U}')^2
			\right]   .
		\end{eqnarray}
		The coefficient $a_{2}$, being proportional to $\tilde{U}'$, and since we are considering a constant potential, vanishes. Therefore, Eq.~(\ref{Generalfourthorderpiequation}) reduces to
		
		\begin{eqnarray}\label{Reducedfourthorderpiequation}
			64 \tilde{G}^{2}\tilde{P}_{\phi}^{4}+ a_{1} 8 \tilde{G}	 \tilde{P}_{\phi}^{2}+a_{3} \approx 0,
		\end{eqnarray}
		whose general solutions are given by
		\begin{eqnarray}
			\tilde{P}_{\phi_{1,2}} &=&  \pm \frac{1}{4\sqrt{\tilde{G}}}\sqrt{-a_{1}-\sqrt{a_{1}^{2}-4a_{3}}}, \nonumber \\
			\tilde{P}_{\phi_{3,4}} &=&  \pm \frac{1}{4\sqrt{\tilde{G}}} \sqrt{-a_{1}+\sqrt{a_{1}^{2}-4a_{3}}}. 
		\end{eqnarray}
		For $a_{3}\ll a_{1}$, the solution $\tilde{P}_{\phi_{3,4}}$ vanishes. Hence, we define the effective Hamiltonian by choosing the positive solution of $\tilde{P}_{\phi_{1,2}}$, i.e.,
		\begin{eqnarray}\label{ReducedHamiltoniana1}
			\mathcal{H}_{\text{eff}}&:=& \tilde{P}_{\phi_{1}}=   \frac{1}{4\sqrt{\tilde{G}}}\sqrt{-a_{1}-\sqrt{a_{1}^{2}-4a_{3}}}. 
		\end{eqnarray}
		Since we are setting the potential to be constant, the internal time fluctiations $G^{X\phi}$ in (\ref{coefficientsa123}) drop, where $X$ is any other field. Now, expanding perturbatively to first order in the potential $\tilde{U}$, the quantum moments, and $\hbar$, straightforward algebra leads to
		\begin{eqnarray}\label{ReducedHamiltoniana2} \mathcal{H}_{\text{eff}} &\approx& -i \frac{\Delta J}{2\sqrt{2\tilde{G}}}-\frac{i\sqrt{2\tilde{G}}\,p^{3}\tilde{U}}{\Delta J} -3i\sqrt{2\tilde{G}}\frac{p\,\tilde{U}}{\Delta J}G^{pp}  -i\sqrt{2\tilde{G}}\frac{p^{3}\tilde{U}}{(\Delta J)^3} \left(G^{JJ}+G^{\bar{J}\bar{J}}-2G^{J\bar{J}}\right) +3i\sqrt{2\tilde{G}} \frac{p^{2}\tilde{U}}{(\Delta J)^2} \left(G^{Jp}-G^{\bar{J}p}\right) \nonumber \\ && -i\frac{\hbar p}{2\sqrt{2\tilde{G}}\Delta J} +\frac{3}{4}i\sqrt{2\tilde{G}} \frac{p^{2}\tilde{U}}{(\Delta J)^2} \left(3\hbar J+\hbar \bar{J}\right) +3i\sqrt{2\tilde{G}} \frac{\hbar p^{4}\tilde{U}}{(\Delta J)^3}, 
		\end{eqnarray}
		where,  the last three terms are operator-ordering contributions. We now set $\sqrt{2\tilde{G}}=\sqrt{2 \gamma^{2}G}= \gamma l$ and rewrite $\tilde{U}=U/2$ to obtain
		\begin{eqnarray}\label{ReducedHamiltoniana3} \mathcal{H}_{\text{eff}} &\approx& -\frac{i}{2l\gamma}\Delta J - \frac{i l \gamma }{2} \frac{p^{3}U}{\Delta J}  -\frac{3 i l \gamma}{2} \frac{p \, U}{\Delta J}  G^{pp} - \frac{i l \gamma}{2} \frac{ p^{3} U}{(\Delta J)^{3}}\left(G^{JJ}+G^{\bar{J}\bar{J}}-2G^{J\bar{J}}\right)   + \frac{3}{2}\frac{i l\gamma Up^{2}}{(\Delta J)^{2}}\left(G^{Jp}-G^{\bar{J}p}\right) +\mathcal{A}, \nonumber \\
		\end{eqnarray}
		
		where $\mathcal{A}$ encodes the last three terms of (\ref{ReducedHamiltoniana2}). Hence, in the absence of scalar perturbations, this equation agrees with Eq.~(\ref{SecondOrderEffectiveHamiltonian}) up to operator-ordering contributions proportional to $\hbar$. 
		
		\section{Explicit form of the effective dynamics.}\label{Appendix}
		The explicit form of the component vectors $F_\mathrm{2nd}$ and
		$G_\mathrm{2nd}$ defined in~(\ref{componentsSecondOrderEOM1}) and~(\ref{componentsSecondOrderEOM2}) are
		given below. 
		Throughout, superscripts $\pm$ on the auxiliary functions $g_i^\pm$
		denote symmetric ($+$) and antisymmetric ($-$) combinations of
		cross-sector moments, fully defined in Eq.~\eqref{eq:aux2}; for example,
		$g_2^+\equiv G^{Jp}+G^{\bar{J}p}$ and
		$g_2^-\equiv G^{Jp}-G^{\bar{J}p}$.
		We use the shorthand $\kappa := \gamma l^2$,
		$\Delta J \equiv J - \bar{J}$, and $\Sigma_J \equiv J + \bar{J}$.
		The auxiliary functions $g_i$ and $b_i$ are defined in~\eqref{eq:aux2}.
		
		The components $F_J$, $F_p$ govern the gravitational background
		evolution and reduce to those of Ref.~\cite{bojowald2007large} when the
		scalar perturbative sector is turned off (i.e., when
		$v_k = \pi_k = 0$ and all cross-sector moments vanish).
		The components $F_{v_k}$, $F_{\pi_k}$, $G_3$, $G_4$,
		and $G_7$--$G_{11}$ constitute the new contributions
		arising from the coupling of the scalar perturbation mode to the
		bouncing background.
		\small{
			\begin{eqnarray}\label{ApendixSecondOrderEOM1}
				F_{J}&=& -\frac{\kappa}{l\gamma}p + \frac{l \gamma \kappa U}{2} \left(\frac{3p^{2}J}{\Delta J}+ \frac{2 p^{4}}{\Delta J^{2}}\right) + l\gamma \kappa H_{\text{s}}\left(\frac{2p^{2}}{\Delta J^{2}}+\frac{J}{\Delta J}\right) + \frac{l \gamma\kappa}{2}\left[\frac{6 U p G^{Jp}}{\Delta J} + \left(\frac{2p^{2}}{\Delta J^{2}} + \frac{J}{\Delta J}\right)g_{3}\right] \nonumber \\
				&&+ \frac{l \gamma\kappa}{2}\left\lbrace 	-\frac{4p}{\Delta J^{3}}b_{2} g_{2}^{-}+g_{1}\left[\frac{2 U p^{2}J}{\Delta J^{3}} + \left(\frac{6 p^{2}}{\Delta J^{4}} + \frac{J}{\Delta J^{3}}\right)b_{2} \right]		 \right\rbrace + l\gamma \kappa \left[\frac{1}{\Delta J}g_{7} + \frac{2p}{\Delta J^{2}}g_{6}\right] \nonumber \\
				&& - l\gamma \kappa \left\lbrace   \left[\frac{4p}{\Delta J^{3}} b_{1}+ \frac{3UpJ}{\Delta J^{2}}\right]g_{2}^{-} + \frac{1}{\Delta J^{2}} b_{1} g_{8}\right\rbrace - l\gamma \kappa \left\lbrace 	-\frac{2p}{\Delta J^{2}}g_{6}+ \left(\frac{4p^{2}}{\Delta J^{3}}+ \frac{J}{\Delta J^{2}}\right)\left[\pi_{k}g_{4} + k^{2}v_{k}g_{5}\right]  \right\rbrace   \nonumber \\
				F_{p} &=& - \frac{\kappa}{2 l \gamma } \left(J+\bar{J}\right) + \frac{l\gamma\kappa p}{2} \frac{\Sigma J}{\Delta J^{2}} \left(Up^{2} +2 H_{\text{s}}\right) + \frac{l\gamma\kappa p }{2}\frac{\Sigma J}{\Delta J^{2}}g_{3} + l\gamma\kappa \frac{\Sigma J}{\Delta J^{2}}g_{6}  + \frac{l \gamma\kappa p }{2}b_{2} \left[-\frac{2}{\Delta J^{3}}g_{9}+ 3\frac{\Sigma J}{\Delta J^{4}}g_{1}\right] \nonumber \\
				&& - l\gamma \kappa b_{1} \left[-\frac{1}{\Delta J^{2}}g_{2}^{+} + 2\frac{\Sigma J}{\Delta J^{3}}g_{2}^{-}\right] - l\gamma\kappa p \left[2\frac{\Sigma J}{\Delta J^{3}}-\frac{1}{\Delta J^{2}}\right]\left[\pi g_{4}+k^{2}v_{k}g_{5}\right]  \nonumber \\
				F_{v_{k}} &=& -\frac{il \gamma p \pi_{k}}{\Delta J} - \frac{i l \gamma p \pi_{k}}{\Delta J^{3}}g_{1} + \frac{i l\gamma \pi_{k}}{\Delta J^{2}}g_{2}^{-}-\frac{il\gamma}{\Delta J}G^{p\pi} + \frac{il\gamma p}{\Delta J^{2}}g_{4}  \nonumber \\
				F_{\pi_{k}} &=& \frac{il\gamma p k^{2}v_{k}}{\Delta J} + \frac{i l\gamma p k^{2}v_{k}}{\Delta J^{3}}g_{1} - \frac{i l\gamma k^{2}v_{k}}{\Delta J^{2}}g_{2}^{-}+\frac{il\gamma k^{2}}{\Delta J}G^{p v} - \frac{il\gamma k^{2} p}{\Delta J^{2}}g_{5}, \nonumber \\
				G_{1}&=& -2\frac{\kappa}{l\gamma}G^{Jp} + \frac{l\gamma \kappa U}{2} \left(\frac{4p^{3}}{\Delta J^{2}}G^{Jp}+\frac{6 p^{2}}{\Delta J}G^{JJ}\right) + l \gamma\kappa H_{\text{s}} \left(\frac{4p}{\Delta J^{2}}G^{Jp}+\frac{2 }{\Delta J}G^{JJ}\right) + 6 \frac{l\gamma \kappa U J p }{\Delta J}G^{Jp} +  2 l\gamma \kappa \left(\frac{2 p^{2}}{\Delta J^{2}} + \frac{J}{\Delta J}\right) g_{7}\nonumber \\
				&&- \frac{2 l\gamma \kappa }{\Delta J^{2}}b_{1} \left[\Sigma JG^{JJ}-4pG^{Jp} \right] - \frac{4 l\gamma \kappa p}{\Delta J^{3}}b_{2} \left[p\left(G^{JJ}-G^{J\bar{J}}-2G^{pp}\right) + JG^{\bar{J}p} + \bar{J}G^{Jp}\right],\nonumber \\
				G_{2} &=& -\left(\frac{\kappa}{l\gamma} - \frac{l\gamma \kappa U}{2}\frac{2p^{3}}{\Delta J^{2}} - l\gamma \kappa H_{\text{s}}\frac{2p}{\Delta J^{2}} \right)g_{2}^{+} + 2 l\gamma \kappa b_{1}\frac{\Sigma J}{\Delta J^{2}} G^{pp}  - 2 l\gamma \kappa p \frac{\Sigma J}{\Delta J^{3}} \left(Up^{2}+2 H_{\text{s}}\right)g_{2}^{-}  + 2 l \gamma \kappa p  \frac{\Sigma J }{\Delta J^{2}}g_{6} \nonumber \\
				G_{3} &=& - \frac{2 i l\gamma p }{\Delta J}G^{v\pi}  + \frac{2 i l\gamma p \pi_{k}}{\Delta J^{2}}g_{5} - \frac{2il\gamma \pi_{k}}{\Delta J}G^{pv} \nonumber \\
				G_{4}&=&  \frac{2 i l\gamma pk^{2} }{\Delta J}G^{v\pi}  - \frac{2 i l\gamma p k^{2}v_{k}}{\Delta J^{2}}g_{4} + \frac{2il\gamma k^{2}v_{k}}{\Delta J}G^{p\pi} \nonumber \\
				G_{5} &=& -\left(\frac{\kappa}{l\gamma} - \frac{l\gamma \kappa U}{2}\frac{2p^{3}}{\Delta J^{2}} - l\gamma \kappa H_{\text{s}}\frac{2p}{\Delta J^{2}} \right)g_{2}^{+} +  \frac{3 l\gamma \kappa U p}{\Delta J} \left(J G^{\bar{J}p}-\bar{J}G^{Jp}\right)   - \frac{l\gamma \kappa}{\Delta J^{2}} b_{1} \left(-2 g_{2}^{+} p  -JG^{\bar{J}\bar{J}}  - \bar{J}G^{JJ} +\Sigma JG^{J\bar{J}} \right)   \nonumber \\
				&& - \frac{2l\gamma \kappa p^{2}}{\Delta J^{3}}b_{2} g_{9} + \frac{2 l\gamma \kappa p^{2}}{\Delta J^{2}} \left[\pi_{k}g_{4}^{+}+ k^{2}v_{k}g_{5}^{+}\right]  + \frac{ l\gamma \kappa }{\Delta J} \left[\pi_{k}\left(JG^{\bar{J}\pi}- \bar{J}G^{J\pi}\right)+ k^{2}v_{k}\left(JG^{\bar{J}v}-\bar{J}G^{Jv}\right)\right] \nonumber \\
				G_{6}&=&   -\frac{\kappa}{2l\gamma}g_{10}+ \frac{l\gamma \kappa U}{2} \left[\frac{p^{3}}{\Delta J^{2}}g_{10} +\frac{3p^{2}G^{Jp}}{\Delta J} \right]  + l\gamma \kappa H_{\text{s}} \left[\frac{p}{\Delta J^{2}}g_{10} +\frac{G^{Jp}}{\Delta J} \right] + 3 l\gamma \kappa U\frac{ p J }{\Delta J}G^{pp} + \frac{2 l\gamma \kappa J}{\Delta J} g_{6}\nonumber \\
				&&  + l\gamma \kappa \frac{p}{\Delta J^{3}} b_{2}\left(2p\left(2G^{\bar{J}p}-G^{Jp}\right)-\Sigma JG^{JJ}-JG^{\bar{J}\bar{J}} + JG^{J\bar{J}} \right) + \frac{l\gamma \kappa}{\Delta J^{2}}b_{1}\left(6p G^{pp}-JG^{\bar{J}p}-\bar{J}G^{Jp}\right)  \nonumber \\
				&&+ \frac{l\gamma \kappa  p}{\Delta J^{2}} \left[\pi_{k}\left((2J+\bar{J})G^{J\pi}+2pG^{p\pi}\right)+k^{2}v_{k}\left((2J+\bar{J})G^{Jv}+2pG^{pv}\right) \right]  \nonumber \\
				G_{7} &=&  - \frac{\kappa}{l\gamma}G^{vp} + \frac{l\gamma \kappa U}{2}\left(\frac{2p^{3}G^{vp}}{\Delta J^{2}} + \frac{3p^{2}G^{Jv}}{\Delta J}\right) + l\gamma \kappa H_{\text{s}}\left(\frac{2 p G^{vp}}{\Delta J^{2}} + \frac{G^{Jv}}{\Delta J}\right)- \frac{il\gamma}{\Delta J}\left[\pi_{k} \left(i\kappa J G^{v\pi}+G^{Jp}\right)+ 2i \kappa k^{2} v_{k}JG^{vv}\right] \nonumber \\
				&&  - \frac{l\gamma \kappa b_{1}}{\Delta J^{2}}\left(J\left(2G^{Jv}-G^{\bar{J}v}\right)-2pG^{pv}\right) + \frac{i l\gamma p}{\Delta J^{2}} \left[\pi_{k}\left(2G^{JJ}-2i\kappa p G^{v\pi}-G^{J\bar{J}}\right)-2 i \kappa k^{2}p v_{k} G^{vv}\right]- \frac{i l\gamma p}{\Delta J}\left(3 i \kappa U J G^{pv} + G^{J\pi}\right) \nonumber \\
				G_{8} &=&  - \frac{\kappa}{l\gamma}G^{p\pi} + \frac{l\gamma \kappa U}{2}\left(\frac{2p^{3}G^{p\pi}}{\Delta J^{2}} + \frac{3p^{2}G^{J \pi}}{\Delta J}\right) + l\gamma \kappa H_{\text{s}}\left(\frac{2 p G^{p\pi}}{\Delta J^{2}} + \frac{G^{J\pi}}{\Delta J}\right)   - \frac{il\gamma}{\Delta J}\left[2 i \kappa \pi_{k}JG^{\pi \pi}+ k^{2} v_{k}\left(i\kappa J G^{v\pi}-G^{Jp}\right)\right]\nonumber \\
				&&  - \frac{l\gamma \kappa b_{1}}{\Delta J^{2}}\left(J\left(2G^{J\pi}-G^{\bar{J}\pi}\right)-2pG^{p\pi}\right) - \frac{i l\gamma p}{\Delta J^{2}} \left[2i \kappa \pi_{k} p G^{\pi \pi}+ k^{2} v_{k} \left(2G^{JJ}+2i\kappa p G^{v\pi}-G^{J\bar{J}}\right)\right] - \frac{i l\gamma p}{\Delta J}\left(3 i \kappa U J G^{p\pi} - G^{Jv}\right) \nonumber \\
				G_{9} &=& -\left(\frac{\kappa}{2 l\gamma} - \frac{l\gamma \kappa}{2}\frac{Up^{3}}{\Delta J^{2}}-\frac{l\gamma \kappa H_{\text{s}}p}{\Delta J^{2}}\right)g_{5}^{+} - \frac{i l\gamma p G^{p\pi}}{\Delta J}  - l\gamma \kappa p \frac{\Sigma J}{\Delta J^{3}}b_{2} g_{5}^{-} - \frac{2 i l\gamma \pi_{k}G^{pp}}{\Delta J} + 2l\gamma \kappa \frac{\Sigma J}{\Delta J^{2}}b_{1}G^{vp}\nonumber \\
				&&  + \frac{i l\gamma p}{\Delta J^{2}}\left[\pi_{k}\left(G^{Jp}-G^{\bar{J}p}-i\kappa \Sigma JG^{v\pi}\right)-2i\kappa k^{2}v_{k}\Sigma JG^{vv}\right]\nonumber \\
				G_{10} &=& -\left(\frac{\kappa}{2 l\gamma} - \frac{l\gamma \kappa}{2}\frac{Up^{3}}{\Delta J^{2}}-\frac{l\gamma \kappa H_{\text{s}}p}{\Delta J^{2}}\right)g_{4}^{+} + \frac{i l\gamma k^{2} p G^{p v}}{\Delta J}  - l\gamma \kappa p \frac{\Sigma J}{\Delta J^{3}}b_{2}g_{4}^{-} + \frac{2 i l\gamma k^{2} v_{k}G^{pp}}{\Delta J} + 2l\gamma \kappa \frac{\Sigma J}{\Delta J^{2}}b_{1}G^{p \pi}\nonumber \\
				&&  + \frac{i l\gamma p}{\Delta J^{2}}\left[-2i \kappa \pi_{k}\Sigma JG^{\pi \pi} +  k^{2}v_{k}\left(G^{\bar{J}p}-G^{Jp}-i\kappa \Sigma JG^{v\pi}\right)\right]\nonumber \\
				G_{11} &=& -\frac{i l\gamma p}{\Delta J}\left(G^{\pi \pi}-k^{2}G^{vv}\right) + \frac{2i l\gamma p}{\Delta J^{2}}\left[\pi_{k}g_{4}^{-} - k^{2}v_{k}g_{5}^{-}\right] - \frac{2 i l\gamma}{\Delta J}g_{6}^{-},
				\label{eq:aux1}
			\end{eqnarray}
		}
		where $\kappa=\gamma l^{2}$, $\Delta J \equiv J-\bar{J}$, $\Sigma J \equiv J +\bar{J}$, and the functions $g_{i}$ and $b_{i}$ are 
		\begin{eqnarray}
			g_{1}=\left(G^{JJ}+G^{\bar{J}\bar{J}}-2G^{J\bar{J}}\right),&\hspace{5cm}&  g_{7}^{\pm}= \left(\pi_{k}G^{J\pi} \pm k^{2}v_{k}G^{Jv}\right), \nonumber \\
			g_{2}^{\pm}= \left(G^{Jp}\pm G^{\bar{J}p}\right), &\hspace{5cm}&  g_{8}=\left(G^{JJ}-G^{\bar{J}\bar{J}}-2G^{pp}\right),\nonumber \\
			g_{3}= \left(3UG^{pp}+G^{\pi \pi }+ k^{2}G^{vv}\right), &\hspace{5cm}& g_{9}= \left(G^{JJ}-G^{\bar{J}\bar{J}}\right) ,\nonumber \\
			g_{4}^{\pm}= \left(G^{J\pi}\pm G^{\bar{J}\pi}\right),&\hspace{2cm}& g_{10}=\left(G^{JJ}+G^{J\bar{J}}+2G^{pp}\right) ,\nonumber \\
			g_{5}^{\pm}  = \left(G^{J v} \pm G^{\bar{J} v}\right), &\hspace{5cm}& b_{1}=\left(\frac{3Up^{2}}{2}+H_{\text{s}}\right),\nonumber \\
			g_{6}^{\pm}= \left(\pi_{k}G^{p\pi}\pm k^{2}v_{k}G^{pv}\right), &\hspace{5cm}& b_{2}= \left(Up^{2}+2H_{\text{s}}\right).
			\label{eq:aux2}
		\end{eqnarray} 
	\end{widetext}
	
	\bibliographystyle{unsrt}
	\bibliography{References}
	
\end{document}